\documentstyle[twoside,12pt,epsf,a4]{article}
\begin{document}
\thispagestyle{empty}
\begin{center}
\begin{LARGE}
\begin{bf}
Black Hole Evaporation. A Survey\\
\end{bf}
\end{LARGE}
\vspace{2.0cm}
Farid Benachenhou\\
\vspace{0.5cm}
Department of Theoretical Physics\\
Uppsala University\\
Box 803\\
S-751 08 Uppsala, Sweden\\
\vspace{2.0cm}
Thesis for the degree of Teknologie Licenciat\\
\vspace{3.0cm}
\begin{abstract}
This thesis is a review of black hole evaporation with emphasis on recent
results obtained for two dimensional black holes. First, the geometry of the
most general stationary black hole in four dimensions is described and some
classical quantities are defined. Then, a derivation of the spectrum of the
radiation emitted during the evaporation is presented. In section four, a two
dimensional model which has black hole solutions is introduced, the so-called
CGHS model. These two dimensional black holes are found to evaporate. Unlike
the four dimensional case, the evaporation process can be studied analytically
as long as the mass of the black hole is well above the two dimensional analog
of the Planck mass. Finally, some proposals for resolving the so-called
information paradox are reviewed and it is concluded that none of them is fully
satisfactory.
\end{abstract}  
\vspace{3cm}
Uppsala 1994
\end{center}
\newpage
\thispagestyle{empty}
\mbox{}
\newpage
\thispagestyle{empty}
\tableofcontents
\newpage
\thispagestyle{empty}
\mbox{}
\setcounter{page}{0}

\newpage
\section{Introduction}
In this work, we will review some of the known classical and quantum properties
of black
holes, with special emphasis on two dimensional black holes.\\\\
The general theory of relativity shows that if a star is sufficiently massive,
no force can
counterbalance its gravitational force so that it will collapse under its
own weight. This theory further predicts that the endpoint of the
collapse is a black hole, which is an object from which nothing can come out,
at least if quantum effects are neglected. Some classical properties of four
dimensional black
holes are reviewed in section 2.\\\\
In 1975, Hawking discovered that black holes are not black: if quantum effects
are taken into account, the black hole is found to emit particles with a thermal
spectrum. In Hawking's calculation, which is described in section 3, the
gravitational field of the black hole is treated classically but the matter
radiated away is assumed to propagate in the fixed classical spacetime geometry
of the black hole by the laws of quantum mechanics.\\\\
Since the outgoing Hawking radiation is a form of energy, it should modify the
gravitational field of the black hole; this effect is called the back reaction
of the radiation on the geometry. In particular, the back reaction should cause
the mass of the black hole to decrease, the mass of the black hole being a
property of the spacetime geometry. The effect of the back reaction is very
difficult to analyze in $3+1$ dimensions; therefore, one is tempted to simplify
the problem by going to $1+1$ dimensions.\\\\
Such a two dimensional model was proposed in 1991 by Callan, Giddings, Harvey
and Strominger. This so-called CGHS model is the subject of section 4.\\\\
At the classical level, the CGHS model has black hole solutions (see section
4.1). Furthermore,
the effect of back reaction can be studied at the (matter) one-loop level
(section 4.3). One
finds that the mass of the black hole decreases, as expected. Indeed, in one
particular version of the CGHS model, the so-called RST model, proposed by
Russo, Susskind and Thorlacius, the black hole is found to evaporate
completely (section 4.3).\\\\
However, the main motivation for studying two dimensional models was to resolve
the problem of information loss: Hawking discovered that the information
contained in the matter which formed the black hole cannot be encoded in the
outgoing radiation because the radiation is purely thermal. Therefore, most of
the information is stored in the black hole. If the black hole evaporates
completely, Hawking's calculation suggests that the information content will be
lost and this implies that the evolution of the black hole is not described by
a unitary $S$-matrix; therefore, quantum mechanics would not apply to black
holes.\\\\
The hope was therefore raised that the two dimensional models would resolve the
issue of information loss but unfortunately, no definite conclusions could be
drawn.\\\\
Several proposals have been made to resolve the information paradox in four
dimensions, but as we
will see in the conclusions, none of them is fully satisfactory today.    

\newpage
\section{Classical Black Holes}
In this section, we describe the geometry of the most general stationary black
hole, the so-called Kerr black hole; we also define some useful quantities
and finally, we derive some of its thermodynamical properties. More detailed
reviews of classical black holes can be found in
\cite[chapters 6 and 12]{Wald} and \cite[chapters 31-34]{grav}.\\\\
The interaction between the gravitational and the electromagnetic field is 
governed by the coupled Einstein-Maxwell equations. In the Lorentz gauge 
$\nabla^{a} A_{a} = 0$, they are:
\[
\left\{
\begin{array}{lcl}
\nabla^{a} \nabla_{a}\, A_{b} - R^{\,\,\,d}_{ b} A_{d} & = & 0 \\
R_{a b} - {1 \over 2}\, g_{a b}\, R & = & 8 \pi\,T_{a b}\;,
\end{array}
\right.
\]\\
where $T_{a b}$ is the electromagnetic stress-energy tensor:
\[
T_{a b} = {1\over4\pi} \{ F_{ac}\,F^{c}_{ b} - 
{1\over4}\,g_{ab}\,F_{de}\,F^{de} \} \;.\]\\\\
The general stationary black hole solution subject to the constraints that the
mass, angular momentum and electric charge of the black hole take definite 
values, is described by the Kerr metric
\begin{eqnarray}
\label{eq:Kerr}
ds^{2} &=& -\left({\Delta-a^2\sin^2\theta\over\Sigma}\right)dt^2 -
 {2a\sin^2\theta\left(r^2+a^2-\Delta\right)\over\Sigma}\,dt\,d\varphi\nonumber\\
& & + \left({{\left(r^2+a^2\right)}^2-
\Delta a^2 \sin^2\theta\over\Sigma}\right)\sin^2\theta\, d\varphi^2
 + {\Sigma\over\Delta}\,dr^2 + \Sigma\,d\theta^2\;,
\end{eqnarray}
and the electromagnetic potential
\begin{equation}
A_{a} = -{Qr\over\Sigma}\left({\left(dt\right)}_a -
 a\sin^2\theta{\left(d\varphi\right)}_a\right)\;,
\end{equation}
where
\begin{equation}
\left\{
\begin{array}{lcl}
\Sigma &=& r^2 + a^2\cos^2\theta\\
\Delta &=& r^2 + a^2 + Q^2 - 2Mr\;. 
\end{array}
\right.
\end{equation}\\\\
$Q$, $M$, $a$ are the three parameters of the family of solutions and can be 
verified to be the electric charge, the mass and the angular momentum per unit
mass of the black hole, respectively.\\\\
When $a=0$, the metric reduces to the Reissner-Nordstrom metric
\begin{eqnarray}
\label{eq:ReiNo}
ds^2  & = & -\left(1-{2M\over r}+{Q^2\over r^2}\right)dt^2  + 
{\left(1-{2M\over r}+{Q^2\over r^2}\right)}^{-1}dr^2 \nonumber \\
& & + r^2\left(d\theta^2+\sin^2\theta\,d\varphi^2\right)\;,
\end{eqnarray}\\
and when $a=Q=0$, we recover the Scharzschild metric
\begin{eqnarray}
ds^2 & = & -\left(1-{2M\over r}\right)dt^2  + 
{\left(1-{2M\over r}\right)}^{-1}dr^2 \nonumber \\
& & + r^2\left(d\theta^2+\sin^2\theta\,d\varphi^2\right)\;,
\end{eqnarray}
which for symplicity we will analyze below.\\\\
We define
\[
r^{*} = r+2M\ln\left({r\over 2M}-1\right)\;,
\]
so that
\[
ds^{2} = -\left(1-{2M\over r}\right)\left(dt^2-dr^{* 2}\right) +
 r^2\left(d\theta^2+\sin^2\theta\,d\varphi^2\right)\;.
\]\\\\            
Note that $r=2M$ and $r=+\infty$ correspond to 
$r^{*}=-\infty$ resp $r^{*}=+\infty$.\\\\  
Introducing the null coordinates $\left\{\begin{array}{c} u\\v 
\end{array}\right\}
= t\mp r^{*}$, we get (suppressing the angular dependence)
\[\begin{array}{lcl}
ds^2 &=& -\left(1-{2M\over r}\right)du\;dv\\\\
&=& -{2M\over r}\,e^{-{r\over 2M}}\;e^{{v-u\over 4M}}du\,dv\;.
\end{array}\]\\\\
Defining the null coordinates
\[
\left\{
\begin{array}{llc}
U = -e^{-{u\over 4M}}& &-\infty<U<0\\\\
V =\ \ e^{{v\over 4M}}& &\ \ \ 0<V<\infty\;,
\end{array}
\right.
\]
the metric takes the form
\[\begin{array}{lll}
ds^2 &=& -{32M^3\over r}\,e^{-{r\over 2M}}\,dU\,dV\nonumber\\\\
&=& {32M^3\over r}\,e^{-{r\over 2M}}\left(-dT^2+dX^2\right)\;,
\end{array}\]
where
\[
\left\{
\begin{array}{c}
U = T-X\nonumber\\
V = T+X\;.
\end{array}
\right.
\]\\\\
The relation between $(t, r)$ and $(T, X)$ is
\[
\left\{
\begin{array}{lllllllll}
X^2-T^2&=&-UV&=&e^{{v-u\over 4M}}&=&e^{{r^{*}\over 2M}}&=&
\left({r\over 2M}-1\right)e^{{r\over 2M}}\nonumber\\\\
\frac{X+T}{X-T}&=&\frac{V}{-U}&=&e^{{u+v\over 4M}}&=&e^{{t\over 2M}}\;,&&
\end{array}
\right.
\]
or
\[
\left({r\over 2M}-1\right)e^{{r\over 2M}}=X^2-T^2\;,
\]
and
\[
{t\over 2M}=2\arctan{T\over X}=\ln \frac{1+{T\over X}}{1-{T\over X}}\;.
\]
\begin{figure}
	\vspace{6 cm}
	\includegraphics{figbh/Sch1.eps}
	\caption{}
	\label{fig:Sch1}
\end{figure}
\begin{figure}
	\vspace{5 cm}
	\includegraphics{figbh/Sch2.eps}
	\caption{}
	\label{fig:Sch2}
\end{figure}
\\\\In these coordinates, the singularity in the metric components at $r=2M$ has
disappeared and as a consequence, the spacetime can be extended by allowing the
ranges of $U$ 
and $V$ to be unrestricted. The resulting maximally extended Scwarzschild
spacetime is depicted in fig.~\ref{fig:Sch1}.\\\\
Another representation of this spacetime can be obtained by mapping $U,V$ to
null coordinates whose ranges are restricted to finite intervals, for example
\[
\left\{
\begin{array}{c}
U = \tan{1\over 2}\tilde{U}\nonumber\\\\
V = \tan{1\over 2}\tilde{V}\;.
\end{array}
\right.
\]\\\\
The resulting "conformal" or "Penrose" diagram is shown in fig.~\ref{fig:Sch2}.
\\\\The most salient feature of the maximally extended Schwarzschild spacetime
is 
that it has a future event horizon $H^{+}$, i.e.\,, a null hypersurface from 
behind which it is impossible to escape to future null infinity $J^{+}$ without
exceeding the speed of light. It also has a past event horizon $H^{-}$, i.e.\,,
a null hypersurface behind which it is impossible to go when
starting from past null infinity $J^{-}$.\\\\
Both the past and the future event horizon are located at $r=2M$, see 
fig.~\ref{fig:Sch1}.\\\\
The singularity at $r=0$ is a true physical singularity--- \,for example,  
$R_{abcd}R^{abcd}$ is infinite there. However, thanks to the horizons, it cannot
be seen from outside the black hole, that is, from region $I$ in
fig.~\ref{fig:Sch1} and fig.~\ref{fig:Sch2}.\\\\
For a collapsing star, only part of region $I$ and $II$ will be generated (see
fig.~\ref{fig:Coll}).
\begin{figure}
	\vspace{6 cm}
	\includegraphics{figbh/Coll.eps}
	\caption{}
	\label{fig:Coll}
\end{figure}
\\\\As for the general Kerr black hole, it has been proved that it is the only 
possible stationary vacuum black hole, i.e.\,, a classical black hole is 
characterized uniquely by its mass, angular momentum and electric charge: a
black hole has no hair. To show this in a special case, we try to find the
stationary solutions of the Klein-Gordon equation in the Schwarzschild metric.
Decomposing the solution in spherical harmonics, we will show later that the
radial part $f_{l,m}\left(r\right)$ satisfies
\begin{equation}
\partial_{r^{*}}^{\,2} f_{l,m}\left(r\right)=\left(1-{2M\over r}\right)
\left(m^2+{2M\over r^3}+\frac{l\left(l+1\right)}{r^2}\right)
f_{l,m}\left(r\right)\;.
\end{equation}
Because the second derivative with respect to $r^{*}$ is strictly positive 
outside the black hole, a solution which vanishes exponentially at infinity must
blow up at the horizon since $r^{*}\rightarrow-\infty$ there. Thus, there is no 
physically acceptable stationary solution to the Klein-Gordon equation, and a 
Schwarzschild black hole has no Klein-Gordon charge.\\\\
From (\ref{eq:Kerr}), we see that the Kerr metric is singular where $\Sigma=
r^2+a^2\cos^2\theta=0$ and where $\Delta=r^2+a^2+Q^2-2Mr=0$.\\
Evaluation of 
curvature invariants such as $R_{abcd}R^{abcd}$ shows that the singularity at
$\Sigma=0$ is a real singularity when $M\neq0$, which cannot be removed by 
extending the manifold.\\
When $Q^2+a^2>M^2$, there are no solutions to the 
equation $\Delta=0$: we are left with a "naked" singularity at $\Sigma=0$, so
in this case the Kerr solution is not physically acceptable.\\
However, when 
$Q^2+a^2\leq M^2$, $\Delta$ vanishes at $r_{\pm}=M\pm\sqrt{M^2-a^2-Q^2}$. It 
has been shown that the latter singularities (at $r=r_{\pm}$) are of the same
nature as the $r=2M$ singularity in the Schwarzschild spacetime.
In particular, the horizon at 
$r=r_{+}$ prevents an external observer from seeing the real singularity at 
$\Sigma=0$.
\begin{figure}
	\vspace{10 cm}
	\includegraphics{figbh/Kerr.eps}
	\caption{}
	\label{fig:Kerr}
\end{figure}
\\\\A conformal diagram of the extended charged Kerr spacetime with
$a\neq0$ is shown in fig.~\ref{fig:Kerr},
for the non-extreme case $Q^2+a^2<M^2$.\\\\
In a physically realistic gravitational collapse, the spacetime is expected to
be
qualitatively similar to that depicted in fig.~\ref{fig:Coll} because of
instabilities associated with the so-called Cauchy horizon at $r=r_{-}$.\\\\
For a rotating black hole, $a\neq0$, the Killing field $\chi^{a}$ which 
is a null generator of the horizon does not coincide with the time translation
Killing field $\xi^{a}={\left(\frac{\partial}{\partial t}\right)}^{a}$. The 
other Killing field available being $\psi^{a}=
{\left(\frac{\partial}{\partial \varphi}\right)}^{a}$, we must have 
\[
\chi^{a}=\xi^{a}+\Omega_{H}\,\psi^{a}\;.
\]\\
So the event horizon is rotating with angular velocity $\Omega_{H}$. It can
be calculated by requiring $\chi^{a}$ to be null on the horizon
\[
\begin{array}{ccccccc}
0=\chi^{a}\chi_{a}&=&\xi^{a}\xi_{a}&+&2\,\xi^{a}\psi_{a}\,
\Omega_{H}&+&\psi^{a}\psi_{a}\,\Omega_H^2\\\\
\ &=&\frac{a^2\sin^2\theta}{\Sigma}&-&\frac{2a\sin^2\theta\,\left(r_{+}^2+a^2
\right)}{\Sigma}\,\Omega_H&+&\frac{{\left(r_{+}^2+a^2\right)}^2}
{\Sigma}\sin^2\theta\,\Omega_H^2\;.
\end{array}
\]\\\\
So
\begin{equation}
{\left(\left(r_{+}^2+a^2\right)\Omega_{H}-a\right)}^2=0
\Leftrightarrow\Omega_{H}=\frac{a}{\left(r_{+}^2+a^2\right)}\;.
\end{equation}
\\\\
Another important quantity is the surface gravity of the black hole $\kappa$.
For a static non-rotating black hole, it can be defined as the force exerted by
a stationary observer at infinity on a stationary particle with unit mass (the
particle can be thought of as being connected to the observer by means of a long
massless string).\\\\
The four-velocity of the particle is
\[u^{a}={\left(-\xi^{a}\xi_{a}\right)}^{-{1\over2}}\,\xi^{a}
\equiv{1\over V}\,\xi^{a}\;,\]
where $V^2=-\xi^{a}\xi_{a}$ is the redshift factor, so the acceleration is
\[\begin{array}{ll}
a^{b}=u^{a}\nabla_{a}u^{b}&={1\over V}\,\xi^{a}\nabla_{a}\,{1\over V}\,\xi^{b}=
{1\over V^2}\,\xi^{a}\nabla_{a}\,\xi^{b}\\\\
&=-{1\over V^2}\,\xi^{a}\nabla^{b}\xi_{a}={1\over2}\,{1\over V^2}\,\nabla^{b}
\left(-\xi^{a}\xi_{a}\right)\\\\
&={1\over2}\,{1\over V^2}\,\nabla^{b}\,V^2=\nabla^{b}\ln V\;,
\end{array}\]
and the magnitude of the local force per unit mass is
\[
\sqrt{a^c a_c}={1\over V}\,\sqrt{\left(\nabla_{a}V\,\nabla^{a}V\right)}\;.
\]\\\\
$a$ goes to infinity as one approaches the horizon because the redshift factor
$V^2=-\xi^{a}\xi_{a}=-g_{tt}$\, goes to zero.\\\\
The force at infinity can be defined
as
\[
a_{\infty}^b=-\nabla^{b}e_{\infty}\;,
\]
where $e_{\infty}$ is the energy per unit mass of the stationary particle as
measured by the stationary observer at infinity
\[
e_{\infty}=-\xi^{a}u_{a}=-\xi^{a}\,{1\over V}\,\xi_{a}=V\;.
\]\\\\
We see that
\[
a_{\infty}=Va=\sqrt{\left(\nabla_{a}V\,\nabla^{a}V\right)}\;.
\]\\\\
For a Schwarzschild black hole
\[
V^2=-\xi^{a}\xi_{a}=-g_{tt}=\left(1-{2M\over r}\right)\;,
\]
and
\begin{eqnarray}
\kappa&=&\lim_{r\rightarrow 2M}a_{\infty}=\lim_{r\rightarrow 2M}\sqrt{{g^{rr}}
{\left(\partial_r V\right)}^2}=\lim_{r\rightarrow 2M}\sqrt{g^{rr}}\partial_r V
\nonumber\\
&=&\lim_{r\rightarrow 2M}{1\over\sqrt{g_{rr}}}\,{1\over2V}\,
\partial_r V^2={1\over2}
\left.\partial_r V^2\right|_{r=2M}={1\over2}\left.{2M\over r^2}\right|_{r=2M}
\nonumber\\
&=&{1\over4M}\;.
\end{eqnarray}\\\\
For a rotating black hole, we define $a^b$ by
\[
a^b=\frac{\chi^{a}\nabla_{a}\chi^{b}}{-\chi^{a}\chi_{a}}=\nabla^{b}\ln V\;,
\]
where now
\[\begin{array}{ll}
V^2=-\chi^{a}\chi_{a}&=-\left(\xi^{a}+\Omega_{H}\psi^{a}\right)
\left(\xi_{a}+\Omega_{H}\psi_{a}\right)\\\\
&=-g_{tt}-2\,\Omega_H g_{t\varphi}-\Omega^2_{H}\,g_{\varphi\varphi}\\\\
&={1\over\Sigma}\left\{\Delta-a^2\sin^2\theta+
2a\sin^2\theta\left(r^2+a^2-\Delta\right)\Omega_H\right.\\\\
&-\left.\left({\left(r^2+a^2\right)}^2-\Delta a^2\sin^2\theta\right)\sin^2\theta
\,\Omega^2_{H}\right\}\;.
\end{array}\]\\\\
It can be shown \cite[chapter 12]{Wald}
that $\kappa$ is constant over the event horizon so we can 
evaluate it at $\theta=0$. Using $\partial_{\theta}V\left(r,\theta=0\right)=0$,
we get
\begin{eqnarray}
\label{eq:kappa}
\kappa&=&\sqrt{g^{rr}{\left(\partial_r V\right)}^2}=\sqrt{\frac{\Delta}{\Sigma}}
\,{1\over2V}\,\partial_r V^2={1\over2}\left.\partial_r\left(\frac{\Delta}{\Sigma}
\right)\right|_{r=r_{+},\theta=0}\nonumber\\
&=&{1\over2}\,\frac{\left.\partial_r\Delta\right|_{r=r_{+}}}{r_+^2+a^2}=
{1\over2}\,\frac{r_{+}-r_{-}}{r_+^2+a^2}\nonumber\\
&=&\frac{\sqrt{M^2-a^2-Q^2}}{2M\left(M+\sqrt{M^2-a^2-Q^2}\right)-Q^2}\;.
\end{eqnarray}
\\$\kappa$ appears in the relation between the Killing parameter $v$ along
the null
generator of the horizon defined by $\chi^a\nabla_a v=1$ and the affine
parameter 
$\lambda$ along the same generator: it can be proved \cite[chapter 12]{Wald} 
that
\[
\lambda\propto e^{\kappa v}\;.
\]\\
This shows that the null coordinate $V$ that we defined earlier when we
analyzed 
the Schwarzschild metric is an affine parameter along the future horizon because
we had $V=e^{{v\over4M}}=e^{\kappa v}$.\\\\
The area $A$ of the event horizon of a Kerr black hole is
\[\begin{array}{ll}
A&=\int_{r=r_{+}}\,d\theta\,d\varphi\,\sqrt{g_{\theta\theta}g_{\varphi\varphi}}
\\\\
&=\int_{r=r_{+}}\,d\theta\,d\varphi\,
\sqrt{\Sigma\,\frac{{\left(r_{+}^2+a^2\right)}^2
\sin^2\theta}{\Sigma}}\\\\
&=4\pi\left(2Mr_{+}-Q^2\right)\;.
\end{array}\]\\\\
Clearly, the area of the event horizon is constant in time. This is a special
case
of the black hole area theorem which Hawking proved in 1971 by using global
properties of spacetime: it states that the total area of all black holes in the
universe cannot decrease, $\delta A\geq0$. The area of a black hole is thus 
analogous to the entropy of a thermodynamical system, and in fact, the area
theorem is sometimes called the second law of black hole dynamics. 
\\\\There is
also an analogous first law which we now derive by studying the variation of
$A$
with respect to $M$ and $J\equiv Ma$
\[{1\over8\pi}\,\delta A=r_+\,\delta M+M\,\delta r_+\;.\]\\
Now
\[\begin{array}{lll}
\delta r_+&=&\delta\left(M+\sqrt{M^2-a^2-Q^2}\right)\\\\
&=&\delta M+{1\over2}\,{\left(M^2-a^2-Q^2\right)}^{-{1\over2}}\,2
\left(M\,\delta M-a\,\delta a\right)\\\\
&=&\delta M+{\left(r_+-M\right)}^{-1}\left(M\,\delta M-a\,\delta a\right)\;,\\\\
a\,\delta a&=&a\,\delta\left(J/M\right)={a\over M}
\,\delta J-{a^2\over M}\,\delta M\;.
\end{array}\]\\
So
\[\begin{array}{ll}
{1\over8\pi}\,\delta A&=\left\{r_++M\left(1+{\left(r_+-M\right)}^{-1}
\left(M^2+a^2\over M\right)\right)\right\}\delta M-\frac{a}{\left(r_+-M\right)}
\,\delta J\\\\
&=\frac{r_+^2+a^2}{r_+-M}\,\delta M-\frac{a}{r_+-M}\,\delta J={1\over\kappa}
\,\delta M-\frac{\Omega_H}{\kappa}\,\delta J\;,
\end{array}\]
where we have used $\kappa=\frac{r_+-M}{r_+^2+a^2}$ and $\Omega_H=\frac{a}
{r_+^2+a^2}$.\\\\
Thus
\begin{equation}
\delta M=\frac{\kappa}{8\pi}\,\delta A+\Omega_H\delta J\;.
\end{equation}
\\\\This suggests that the laws of black hole dynamics are really the laws of 
thermodynamics applied to black holes. In this case, the temperature and entropy
of the black hole are $T=\alpha\kappa$ and $S={1\over{8\pi\alpha}}\,A$,
respectively. 
However, 
for classical black holes, the temperature is zero because a classical black
hole 
only absorbs energy without emitting anything. This would be in contradiction
with the conjecture made above. When quantum effects are taken into account, 
this paradox is resolved because, as was first shown by Hawking, a black hole
emits radiation like a blackbody at temperature $\hbar\kappa/2\pi$ 
(see next section), and therefore, the entropy of a black hole is
$S=\hbar^{-1}A/4$.

\newpage
\section{Hawking Radiation}
The black hole radiance was found \cite{Hawk75}
by studying the propagation of quantized,
noninteracting matter fields in the classical geometry of the black hole.\\\\
The simplest example one would consider is a neutral spin zero particle
described by the real Klein-Gordon field which propagates in region I of the
extended Schwarzschild spacetime, fig.~\ref{fig:Sch2}.\\\\
It satisfies the equation 
\[\Box\phi-m^2\phi=0\;,\] 
where
\[
\Box\phi=g^{ab}\nabla_a\nabla_b\phi={\left(-g\right)}^{-{1\over2}}\partial_a
\left({\left(-g\right)}^{1\over2}g^{ab}\partial_b\phi\right)\;.
\]\\\\
Here $-g_{tt}=g_{r^*r^*}=\left(1-{2M\over r}\right)$, $g_{\theta\theta}=r^2$,
$g_{\varphi\varphi}=r^2\sin^2\theta$ and all non-diagonal components are
zero.\\
Also
$\sqrt{-g}=\left(1-{2M\over r}\right)r^2\sin\theta$, so 
\[\begin{array}{lll}
0=\left(\Box-m^2\right)\phi&=&g^{00}\partial_0^2\phi+{\left(-g\right)}^
{-{1\over2}}\partial_{r^{*}}\left({\left(-g\right)}^{1\over2}g^{r^{*}r^{*}}
\partial_{r^{*}}\phi\right)\\\\
&+&{1\over r^2}\left({1\over\sin\theta}\,\partial_{\theta}\left(\sin\theta\,
\partial_\theta\phi\right)+{1\over\sin^2\theta}\,\partial_{\varphi}^2\phi\right)
-m^2\phi\;.
\end{array}\]\\\\
Setting $\phi={1\over r}f\left(r,t\right)Y_{lm}\left(\theta,\varphi\right)$,
the equation becomes:
\[\begin{array}{l}
-{\left(1-{2M\over r}\right)}^{-1}\partial_t^{\,2}{f\over r}+
{\left(1-{2M\over r}\right)}^{-1}r^{-2}\partial_{r^*}\left(\left
(1-{2M\over r}\right)r^2{\left(1-{2M\over r}\right)}^{-1}\partial_{r^*}
{f\over r}\right)\\\\
-\frac{l\left(l+1\right)}{r^2}{f\over r}-m^2{f\over r}=0\;,
\end{array}\]
or
\[\partial_t^2f-r^{-1}\partial_{r^*}\left(r^2\partial_{r^*}{f\over r}\right)+
\left(1-{2M\over r}\right)\left(\frac{l\left(l+1\right)}{r^2}+m^2\right)f=0\;.\]
\\\\
Now 
\[\begin{array}{ll}
\ &\partial_{r^*}\left(r^2\partial_{r^*}{f\over r}\right)=\partial_{r^*}\left(r
\partial_{r^*}f+r^2f\partial_{r^*}{1\over r}\right)\\\\
=&r\partial_{r^*}^2 f+\partial_{r^*}\left(r^2\partial_{r^*}{1\over r}\right)f=
r\partial_{r^*}^2 f+\left(1-{2M\over r}\right)\partial_r\left(r^2
\left(1-{2M\over r}\right){{-1}\over r^2}\right)f\\\\
=&r\partial_{r^*}^2 f-\left(1-{2M\over r}\right){2M\over r^2}f\;,
\end{array}\]
so we end up with the equation:
\begin{equation}
\partial_t^2f-\partial_{r^{*}}^2 f+\left(1-{2M\over r}\right)\left(
\frac{l\left(l+1\right)}{r^2}+{2M\over r^3}+m^2\right)f=0\;.
\end{equation}\\\\
This equation has the form of a wave equation for a massless scalar field in 
a flat two dimensional spacetime with a scalar potential
\[V\left(r^*\right)=\left(1-{2M\over r}\right)\left(\frac{l\left(l+1\right)}
{r^2}+{2M\over r^3}\right)\;,\]
where we have specialized to $m=0$.\\\\
As $r^*\rightarrow-\infty$ or $r\rightarrow2M$, $V\left(r^*\right)\sim
\left(1-{2M\over r}\right)\sim e^{{r^*\over2M}}$, i.e.\,, the potential vanishes
 exponentially.\\
As $r^*\rightarrow+\infty$ or $r\rightarrow+\infty$, the potential vanishes at 
least as ${1\over r^{*2}}$\,.\\\\
Thus, in the asymptotic past, any solution reduces to the sum of two free wave 
packets, one coming from the white hole horizon $H^{-}$, called 
$f_{-}\left(u\right)$, and one coming from past null infinity $J^{-}$,
called $g_{-}\left(v\right)$.\\\\
Similarly, in the asymptotic future, any solution reduces to the sum of two
free wave packets: one going towards the black hole horizon $H^{+}$, 
$g_{+}\left(v\right)$, and one going towards future null infinity $J^{+}$,
$f_{+}\left(u\right)$.\\\\
Any incoming one particle state is thus the linear combination of a one 
particle state coming from $H^{-}$
and a one particle state coming from $J^{-}$
\[\cal H_{in}=\cal H_{H^-}\oplus \cal H_{J^-}\;.\]\\    
$\cal H_{J^-}$ is defined as the vector space of all solutions 
$g_{-}\left(v\right)$ containing only positive frequencies with respect to some 
null coordinate $\tilde v\left(v\right)$
\[g^{\left(+\right)}_-\left(\tilde v\right)={1\over\sqrt{2\pi}}\int_{0}^
{\infty}d\omega\;e^{-i\omega\tilde v}\hat g^{\left(+\right)}_-\left(\omega
\right)\;.\]\\\\
Here $\tilde v=v=t+r^*$ is the natural choice because $J^{-}$ is an
asymptotically flat region and in this region, $\left(u, v\right)$ are
minkowskian (null) coordinates, $ds^2=-du\,dv$.
\\\\However, there is an ambiguity in
the definition of $\cal H_{H^-}$ because $H^-$ is {\em not} asymptotically flat
and as a consequence, there is no unique choice of $\tilde u\left(u\right)$: 
we could choose it to be either the Killing parameter $u=t-r^*$ or the affine
parameter $U=-e^{-{u\over4M}}$.\\\\
For the collapsing spherical body of 
fig.~\ref{fig:Coll}, the spacetime contains no white hole horizon so 
$\cal H_{in}$ is just $\cal H_{J^-}$ and is therefore uniquely defined.\\\\
However,
an ambiguity still remains in the definition of 
$\cal H_{out}=\cal H_{H^+}\oplus \cal H_{J^+}$ because the definition of
positive frequency solutions on the black hole horizon is not unique.\\\\
We assume that the Fock space $\cal F$ of the quantum field $\hat{\phi}$ is
isomorphic
to the Fock space  $\cal F_{in}\left(\cal H_{in}\right)$ of incoming particles, 
where  $\cal F_{in}\left(\cal H_{in}\right)$ is the direct sum of the Hilbert 
spaces of no particle states, one particle states, two particle states and so
on. Let $U:\cal F\rightarrow\cal F_{in}$ denote this isomorphism.\\\\
If we define
\[U\hat{\phi}\left(x\right)U^{-1}=\sum_{i=1}^{\infty}\sigma_i\left(x\right)
a_{in}\left(\overline{\sigma}_i\right)+\overline{\sigma}_i\left(x\right)a_{in}
^{\dagger}\left(\sigma_i\right)\;,\]
where $\left\{\sigma_i\left(x\right)\right\}$ is an orthonormal basis of 
$\cal H_{in}$ for $x$ in the asymptotic past $J^-$, we see that for these
$x$'s,
$U\hat{\phi}\left(x\right)U^{-1}$ is the usual free field operator in flat
spacetime.\\\\
In the same way, we denote by $W$ the isomorphism between  $\cal F$ and
$\cal F_{out}$ and we define
\[W\hat{\phi}\left(x\right)W^{-1}=\sum_{i=1}^{\infty}\rho_i\left(x\right)
a_{out}\left(\overline{\rho}_i\right)+\overline{\rho}_i\left(x\right)a_{out}
^{\dagger}\left(\rho_i\right)\;,\]
where $\left\{\rho_i\left(x\right)\right\}$ is an orthonormal basis of 
$\cal H_{out}$ for $x$ in the asymptotic future.
$W\hat{\phi}\left(x\right)W^{-1}$ then reduces to the free field operator 
in flat spacetime.\\\\
So
\begin{eqnarray}
\label{eq:iso}
&\ &S\left(\sum_{j=1}^{\infty}\sigma_j\, a_{in}\left(\overline{\sigma}_j\right)+
\overline{\sigma}_j\, a_{in}^{\dagger}\left(\sigma_j\right)\right)S^{-1}
\nonumber\\
&=&\sum_{j=1}^{\infty}\rho_j\, a_{out}\left(\overline{\rho}_j\right)+
\overline{\rho}_j\, a_{out}^{\dagger}\left(\rho_j\right)\;,
\end{eqnarray}
where $S=WU^{-1}$.\\\\
Taking the Klein-Gordon 
inner product \cite{Wald75} of this equation with $\sigma_i$,
the left hand side becomes
\[\begin{array}{lll}
&\ &\left(\sigma_i,\sum_{j=1}^{\infty}\sigma_j\,
Sa_{in}\left(\overline{\sigma}_j\right)S^{-1}+\overline{\sigma}_j\,
Sa_{in}^{\dagger}\left(\sigma_j\right)S^{-1}\right)_{K.G.}\\\\
&=&Sa_{in}\left(\overline{\sigma}_i\right)S^{-1}\;.
\end{array}\]\\\\
In the asymptotic future, $\sigma_i\left(x\right)$ reduces to a solution of the
Klein-Gordon equation in flat spacetime and contains both a positive and a 
negative frequency part, denoted by $\mu$ and $\lambda$, respectively.\\\\
Define 
the operators $C:\cal H_{in}\rightarrow\cal H_{out}$ and
$D:\cal H_{in}\rightarrow\cal {\overline {H}}_{out}$ in such a way that
$C\sigma=\mu$ and $D\sigma=\lambda$.\\\\
The Klein-Gordon product of the right hand side of
~(\ref{eq:iso}) with $\sigma_i\left(x\right)$ is
\[\begin{array}{lll}
&\ &\sum_{i=1}^{\infty}{\left(\sigma_i,\rho_j\right)}_{K.G.}
a_{out}\left(\overline{\rho}_j\right)+
{\left(\sigma_i,\overline{\rho}_j\right)}_{K.G.}
a_{out}^{\dagger}\left(\rho_j\right)\\\\
&=&a_{out}\left(\overline{C\sigma}_i\right)-
a_{out}^{\dagger}\left(\overline{D\sigma}_i\right)\;,
\end{array}\]
so we have
\begin{equation}
\label{eq:Smat}
Sa_{in}\left(\overline{\sigma}_i\right)S^{-1}=
a_{out}\left(\overline{C\sigma}_i\right)-
a_{out}^{\dagger}\left(\overline{D\sigma}_i\right)\;.
\end{equation}\\\\
This relation defines the action of the S-matrix. It can be shown that S is
unitary.\\\\
Since $\cal F_{out}\left(\cal H_{H^+}\oplus\cal H_{J^+}\right)$ is isomorphic
to $\cal F_{H^+}\left(\cal H_{H^+}\right)\otimes
\cal F_{J^+}\left(\cal H_{J^+}\right)$ , any  $\left|\psi\right>\in\cal F_{out}$
can be written as:
\[\left|\psi\right>=c_{hj}\left|h\right>\left|j\right>\;,\]
where $c_{hj}=\left<h,j|\psi\right>$, and $\left|h\right>$ and $\left|j\right>$
are orthonormal bases in $\cal F_{H^+}$ resp. $\cal F_{J^+}$.\\\\
If O is an operator acting on $\cal F_{J^+}$, its expectation value in the
state $\left|\psi\right>$ is
\[\begin{array}{lll}
\left<\psi|O|\psi\right>&=&\left<h',j'\right|\overline{c}_{h'j'}\,O\,{c}_{hj}
\left|h,j\right>={c}_{hj}\,\overline{c}_{h'j'}\left<h'|h\right>
\left<j'\right|O\left|j\right>\\\\
&=&{c}_{hj}\,\overline{c}_{hj'}\left<j'\right|O\left|j\right>=
tr_{\cal F_{J^+}}\left\{{c}_{hj}\,\overline{c}_{hj'}
\left|j\right>\left<j'\right|O\right\}\;.
\end{array}\]\\\\
Therefore, a pure state $\left|\psi\right>$ in $\cal F_{H^+}\otimes\cal F_{J^+}$
is viewed as a mixed state in $\cal F_{J^+}$ described by the density matrix
$\rho=\left<h,j|\psi\right>\left<\psi|h,j'\right>\left|j\right>\left<j'\right|$
\begin{equation}
\left<\psi|O|\psi\right>=tr_{\cal F_{J^+}}\left\{\rho O\right\}\;.
\end{equation}\\
A redefinition of the notion of positive frequency on the black hole horizon
induces a \underline{unitary} transformation, of the form given by
~(\ref{eq:Smat}), acting on $\cal F_{H^+}$
\[\left|\psi '\right>=S\left|\psi\right>=c_{hj}\left(S\left|h\right>\right)
\left|j\right>\;.\]\\\\
But since $\left<h'\right|S^{\dagger}S\left|h\right>=\left<h'|h\right>$, 
$\rho$ is left unchanged which means that at $J^{+}$, one obtains unambiguous
physical predictions: the results of measurements performed at $J^{+}$ do not
depend on the choice of positive frequency on the black hole horizon.\\\\
Coming back to the fundamental equation ~(\ref{eq:Smat}), 
\[Sa_{in}\left(\overline{\sigma}_i\right)S^{-1}=
a_{out}\left(\overline{C\sigma}_i\right)-
a_{out}^{\dagger}\left(\overline{D\sigma}_i\right)\;,\]
we can solve for $\left|\psi\right>\equiv S\left|0_{in}\right>$,
where $\left|0_{in}\right>$ is the vacuum state of $\cal F_{in}$: we write
\[\psi=\left(c,\eta^{a_1},\eta^{a_1a_2},\eta^{a_1a_2a_3},...\right)\;,\]
where $\eta^{a_1...a_n}$ is an $n$-particle state of
$\cal F_{out}\left(\cal H_{H^+}\oplus\cal H_{J^+}\right)$, i.e.,
\[\eta^{a_1...a_n}\in \bigotimes_S^n
\left(\cal H_{H^+}\oplus\cal H_{J^+}\right)\;,\]
and apply both sides of 
~(\ref{eq:Smat}) to $\left|\psi\right>$
\[Sa_{in}\left(\overline{\sigma}_i\right)\left|0_{in}\right>=0=
\left\{a_{out}\left(\overline{C\sigma}_i\right)-
a_{out}^{\dagger}\left(\overline{D\sigma}_i\right)\right\}\left|\psi\right>\;,\]
or
\begin{equation}
0=\left(a_{out}\left(\overline{\tau}_i\right)-
a_{out}^{\dagger}\left(E\overline{\tau}_i\right)\right)\left|\psi\right>\;,
\end{equation}
where we have defined $\tau_i\equiv C\sigma_i$ and $E=\overline D\,\overline
C^{-1}$.\\\\
Using the definition of $a_{out}$ and $a^{\dagger}_{out}$, we obtain the 
equations \cite{Wald75}
\[\left\{\begin{array}{rcl}
\eta^a\,\overline \tau_a&=&0\\
\sqrt{2}\,\eta^{ab}\,\overline\tau_a&=&c\,{\left(E\overline\tau\right)}^b\\
\sqrt{3}\,\eta^{abc}\,\overline\tau_a&=&\sqrt{2}\,
{\left(E\overline\tau\right)}^{\left(b\right.}\eta^{\left.c\right)}\\
&.&\\
&.&\\
&.&
\end{array}\right.
,\ \ \  \forall\tau^a\;.\]\\\\
The solution is 
\[\left\{\begin{array}{rclcl}
\eta^{a_1...a_n}&=&0&,&n\ odd\\
\eta^{a_1...a_n}&=&c\,\frac{{\left(n!\right)}^{1\over2}}
{2^{n\over2}\left({n\over2}\right)!}\,\varepsilon^{\left(a_1a_2\right.}
\varepsilon^{a_3a_4}...\varepsilon^{\left.a_{n-1}a_n\right)}&,&
n\ even\;,
\end{array}\right.\]
so 
\begin{equation}
\label{eq:psi}
\psi=\psi\left(\varepsilon^{ab}\right)=c\left(1,0,\sqrt{{1\over2}}
\varepsilon^{ab},0,\sqrt{{3.1\over4.2}}\varepsilon^{\left(ab\right.}
\varepsilon^{\left.cd\right)},0,...\right)\;,
\end{equation}
where $\varepsilon^{ab}$ is the two particle state corresponding to $E$.\\\\
We observe that particle creation occurs only if $D\neq0$ and that the particles
are produced in pairs.\\\\
Let $P_{\omega lm}$ denote the solutions which at $J^+$ have the form,
\[{1\over\sqrt{\omega}}\frac{e^{-i\omega u}}{r}
Y_{l,m}\left(\theta,\varphi\right)\;,\]
and construct the wave packets
\[P_{jnlm}={1\over\sqrt{E}}\int_{jE}^{\left(j+1\right)E}d\omega\,P_{\omega lm}\,
e^{\frac{2\pi in\omega}{E}}\;,\]
where $0<E\ll1$.\\\\
These wave packets are made up of frequencies around $\omega=
\left(j+{1\over2}\right)E$ and are peaked around $u={{2\pi n}\over E}$ with a 
time spread $\Delta u={{2\pi}\over E}$.\\\\
At $J^+$, they  provide an orthonormal basis for
$\cal H_{J^+}$ and as elements of  $\cal H_{J^+}$,
we denote them by $_i\rho^a$, where $i$ stands for $jnlm$.\\\\
For an eternal black hole (not formed by gravitational collapse,
~fig.~\ref{fig:Sch2}), we similarly define the wave packets $Q_{jnlm}$ from the
solutions that reduce to
\[{1\over\sqrt{\omega}}\frac{e^{-i\omega v}}{r}
Y_{l,m}\left(\theta,\varphi\right)\;,\] 
at $H^+$.\\\\
In the case of a black hole formed by gravitational collapse,
~fig.~\ref{fig:Coll}, $v$ is defined such that it agrees with the Killing 
parameter
outside the collapsing matter. The ambiguity in defining  $v$ will not affect
$Q_{jnlm}$ for late times, i.e.\,, for large $n$.\\\\
At $H^+$, we choose the $Q_{jnlm}:s$ as part of our basis in
$\cal H_{H^+}$ and denote them by $_i\sigma^a$.\\\\
Together $\left\{_i\rho^a\right\}$ and $\left\{_i\sigma^a\right\}$ form 
a late time basis of $\cal H_{H^+}\oplus\cal H_{J^+}$.\\\\
Another late time basis in $\cal H_{H^+}\oplus\cal H_{J^+}$ can be constructed
as follows: denote by $X_{\omega lm}$ and $Y_{\omega lm}$ the solutions in the
eternal black hole spacetime that have the form ${1\over\sqrt{\omega}}
\frac{e^{-i\omega u}}{r}Y_{l,m}\left(\theta,\varphi\right)$ at $H^-$ resp.
${1\over\sqrt{\omega}}\frac{e^{-i\omega v}}{r}
Y_{l,m}\left(\theta,\varphi\right)$ at $J^-$, and
form the corresponding wave packets, $X_{jnlm}$ and $Y_{jnlm}$. We must have
\[X_{jnlm}=t_{lm}\left(\omega\right)P_{jnlm}+r_{lm}\left(\omega\right)Q_{jnlm}
\;,\]
and
\[Y_{jnlm}=T_{lm}\left(\omega\right)Q_{jnlm}+R_{lm}\left(\omega\right)P_{jnlm}
\;,\]
where $t$, $T$, and $r$, $R$ are transmission resp. reflection amplitudes.\\\\
The $X_{jnlm}:s$
and the $Y_{jnlm}:s$ are thus elements of a new late time basis in 
$\cal H_{H^+}\oplus\cal H_{J^+}$ and we will call them $_i\lambda^a$ resp.
$_i\gamma^a$. So
\[_i\lambda^a=t_i\;_i\rho^a+r_i\;_i\sigma^a\;,\]
and 
\[_i\gamma^a=T_i\;_i\sigma^a+R_i\;_i\rho^a\;.\]\\\\
$\varepsilon^{ab}$ is determined by the action of the operator $DC^{-1}$ on 
$\left\{_i\lambda^a,\;_i\gamma^a\right\}$: if we propagate the wave packet
corresponding to $_i\gamma^a$ backwards in time from $H^+\cup J^+$, it will be
almost entirely scattered back to $J^-$ by the static Schwarzschild geometry
and the resulting wave packet will be the purely positive frequency wave packet
$Y_{jnlm}$. Hence
\[DC^{-1}{_i\gamma^a}=0\;.\]\\\\
On the other hand, the wave packet corresponding to $_i\lambda ^a$
is completely transmitted (backwards in time) to the surface of the collapsing
body because in the extended Schwarzschild spacetime it would, by definition,
be transmitted in its entirety to $H^{-}$. For an observer on the collapsing
body, the wave packets with sufficiently high $n$ will have the form
\[X_{jnlm}={1\over\sqrt{E}}\int_{jE}^{\left(j+1\right)E}d\omega\,
\left\{{1\over\sqrt{\omega}}\frac{e^{-i\omega
u}}{r}Y_{l,m}\left(\theta,\phi\right)\right\}
e^{\frac{2\pi in\omega}{E}}\;.\]\\\\
Because $u$ goes to infinity on 
the event horizon, the surfaces of constant phase of 
$\phi\left(u\right)\equiv e^{-i\omega u}$ pile up close to it (see
fig.~\ref{fig:mode}). To the observer, the frequency thus appears to go to 
infinity
as the radius of the body goes to $2M$.
\begin{figure}
	\vspace{6 cm}
	\includegraphics{figbh/Mode.eps}
	\caption{}
	\label{fig:mode}
\end{figure}
\\\\Indeed, the dependence of the phase 
on the proper time (or affine parameter) $\lambda$ along the geodesic of the
observer or any other observer entering the black hole, is
\[+\omega u\left(U\left(\lambda\right)\right)=
-{\omega\over\kappa}\ln\left(-U\left(\lambda\right)\right)\;.\]\\\\
Since $U$ depends smoothly on $\lambda$ and satisfies
$\frac{dU}{d\lambda}\neq0$, we can write
\[U\left(\lambda\right)=\left(\frac{dU}{d\lambda}\right)_{\!\!\!\lambda=0}\!\!\!
\lambda\hspace{1cm} \mbox{, close to}\;H^+\;,\]
where $\lambda$ is chosen to be zero at $H^+$. Therefore, close to  the horizon
\[\omega u=-{\omega\over\kappa}\ln\left(-\alpha\lambda\right)\;,\]
where $\alpha\equiv\left(\frac{dU}{d\lambda}\right)_{\lambda=0}$.\\\\
Since the local frequency is very high, for large $u$, $\phi$ will propagate
through the collapsing body and out to $J^-$ by geometric optics.
The null geodesic generators of the surfaces of constant phase that have a 
large 
$u$ when they enter the body will have a $v$ less then but infinitesimally
close to $v=v_0\equiv0$ (where $v_0$ is the continuation backwards in time of 
the event horizon (which has $u=+\infty$)) as they pass the center
(see fig.~\ref{fig:mode}).\\\\
Introducing a geodesic deviation vector $\eta^a$ between these generators 
and choosing its direction at $J^-$ to be along
${\left(\frac{\partial}{\partial v}\right)}^a$, we realize that near 
$v=0$, the $v$-dependence of $\phi$ at $J^-$ will be the same as the dependence
of $\phi\left(\lambda\right)$ on the affine parameter $\lambda$ along the
geodesic tangent to $\eta^a$, for points close to $H^+$:
\begin{equation}
\phi\left(u\left(\lambda\leftarrow v\right)\right)\sim\left\{
\begin{array}{lcl}
0&,&v>0\\
e^{\frac{i\omega}{\kappa}\ln\left(-\alpha v\right)}&,&v<0\;.
\end{array}\right.
\end{equation}\\\\
The wave packet corresponding to $\phi\left(v\right)$ is
\[\begin{array}{rcl}
Z_{jnlm}\left(v\right)&\equiv&{1\over\sqrt{E}}\int_{jE}^{\left(j+1\right)E}
d\omega\,\left\{{1\over\sqrt{\omega}}\frac{\phi\left(v\right)}{r}
Y_{l,m}\left(\theta,\phi\right)\right\}
e^{\frac{2\pi in\omega}{E}}\\\\
&\sim&\left\{\begin{array}{lcl}
0&,&v>0\\
{1\over L}\,e^{{{i\omega_j L}\over E}}\,
\sin{\left({L\over2}\right)}&,&v<0\;,
\end{array}\right.
\end{array}\] 
where $\omega_j=\left(j+{1\over2}\right)E$ is the original frequency of
$X_{jnlm}$ and $L=2\pi n+{E\over\kappa}\ln\left(-\alpha v\right)$. In the last
line, we
have also suppressed all $v$-independent factors.\\\\
The crucial point is
that $Z_{jnlm}$ is not purely positive frequency. In fact, it can be shown
\cite[Appendix A]{Wald75} that
its Fourier transform $\hat Z_{jnlm}\left(\omega'\right)$ satisfies
\begin{equation}
\hat Z_{jnlm}\left(-\omega'\right)=-e^{-{{\pi\omega_j}\over\kappa}}
\hat Z_{jnlm}\left(\omega'\right)\hspace{1cm},\ \omega'>0\;.
\end{equation}\\\\
So
\[\begin{array}{lcl}
Z_{jnlm}\left(v\right)&=&Z^{\left(+\right)}_{jnlm}\left(v\right)+
Z^{\left(-\right)}_{jnlm}\left(v\right)\\\\
\ &=&\int_{0}^{+\infty}d\omega'\,\frac{e^{-i\omega'v}}{\sqrt{\omega'}}
\hat Z_{jnlm}\left(\omega'\right)\\\\
\ &+&\int_{0}^{+\infty}d\omega'\,\frac{e^{i\omega'v}}{\sqrt{\omega'}}
\hat Z_{jnlm}\left(\omega'\right)
\left(-e^{-{{\pi\omega_j}\over\kappa}}\right)\;.
\end{array}\]\\\\
We introduce the time reflected wave packet $\tilde {Z}_{jnlm}\left(v\right)$ at
$J^-$
\[\begin{array}{lcl}
\tilde {Z}_{jnlm}\left(v\right)\equiv Z_{jnlm}\left(-v\right)&=&
\int_{0}^{+\infty}d\omega'\,\frac{e^{-i\omega'v}}{\sqrt{\omega'}}
\hat Z_{jnlm}\left(\omega'\right)
\left(-e^{-{{\pi\omega_j}\over\kappa}}\right)\\\\
\ &+&\int_{0}^{+\infty}d\omega'\,\frac{e^{i\omega'v}}{\sqrt{\omega'}}
\hat Z_{jnlm}\left(\omega'\right)\;.
\end{array}\]\\\\
The $\tilde {Z}_{jnlm}:s$ are orthonormal with negative unit
Klein-Gordon norm, and they are obviously orthogonal to the
$\left\{Z_{jnlm}\right\}$. $\tilde {Z}_{jnlm}\left(v\right)$ will also propagate
by geometric optics since its effective frequency is as high as
$Z_{jnlm}\left(v\right)$: it will end up at the horizon just after its
formation, see fig.~\ref{fig:mode}. The resulting wave packet
$J_{jnlm}\left(v\right)$ we choose to be part of our basis in 
$\overline {\cal {H}}_{H^+}$ and we denote them by
$_i\overline\tau_a$\,. $_i\tau^a
\in \cal {H}_{H^+}$ are thus early time horizon states.\\\\
We note that the combinations
\[\begin{array}{cl}
\ &Z_{jnlm}\left(v\right)+
e^{-{{\pi\omega_j}\over\kappa}}\tilde {Z}_{jnlm}\left(v\right)\\\\
=&\int_{0}^{+\infty}d\omega'\,\frac{e^{-i\omega'v}}{\sqrt{\omega'}}
\hat Z_{jnlm}\left(\omega'\right)
\left(1-e^{-{{2\pi\omega_j}\over\kappa}}\right)\;,
\end{array}\]
and
\[\begin{array}{cl}
\ &Z_{jnlm}^*\left(v\right)+
e^{{{\pi\omega_j}\over\kappa}}\tilde {Z}_{jnlm}^*\left(v\right)\\\\
=&\int_{0}^{+\infty}d\omega'\,\frac{e^{-i\omega'v}}{\sqrt{\omega'}}
\hat Z_{jnlm}^*\left(\omega'\right)
\left(e^{{{\pi\omega_j}\over\kappa}}-
e^{-{{\pi\omega_j}\over\kappa}}\right)\;,
\end{array}\] 
are purely positive frequency wave packets at $J^-$.\\\\
This implies that
\[\left\{\begin{array}{lcl}
D{C^{-1}}_i\lambda^a&=&{e^{-{{\pi\omega_i}\over\kappa}}}_i\overline {\tau}_a\\
DC^{-1}\left({e^{{{\pi\omega_i}\over\kappa}}}_i\tau^a\right)
&=&_i\overline {\lambda}_a\;.
\end{array}\right.\]\\\\
The action of $E=\overline D\,{\overline C}^{-1}$ on the basis elements 
$_i{\overline\gamma}_a$, $_i{\overline\lambda}_a$, $_i{\overline\tau}_a$ in
${\overline{\cal H}}_{H^+}\oplus{\overline{\cal H}}_{J^+}$ is therefore
\[\left\{\begin{array}{ccc}
E\,_i{\overline\gamma}_a&=&0\\\\
E\,_i{\overline\lambda}_a&=&
{e^{-{{\pi\omega_i}\over\kappa}}}_i{\tau}^a\\\\
E\,_i{\overline\tau}_a&=&
{e^{-{{\pi\omega_i}\over\kappa}}}_i{\lambda}^a\;.
\end{array}\right.\]\\\\
Hence $\varepsilon^{ab}$ is
\[\varepsilon^{ab}=\sum_{i}e^{-{{\pi\omega_i}\over\kappa}}\,2\,
_i{\lambda^{\left(a\right.}}_i\tau^{\left.b\right)}+\varepsilon^{ab}_0\;,\]
where $\varepsilon^{ab}_0$ is orthogonal to all the late time basis vectors
$\left\{_i\lambda^a\right\}$ and $\left\{_i\gamma^a\right\}$ as well as the 
early time horizon states $\left\{_i\tau^a\right\}$.\\\\
The state vector $\left|\psi\right>=S\left|0_{in}\right>$ corresponding to the
in-vacuum was given by~(\ref{eq:psi})
\[\psi=\psi\left(\varepsilon^{ab}\right)=\left(1,0,\sqrt{{1\over2}}
\varepsilon^{ab},0,\sqrt{{3.1\over4.2}}\varepsilon^{\left(ab\right.}  
\varepsilon^{\left.cd\right)},0,...\right)\;.\]\\\\
Under the isomorphism between
\[\cal F_{out}\left\{\left(\oplus_i\cal H_i\right)\bigoplus
\left(\oplus_k\cal H_k\right)\bigoplus
\cal H\left(\varepsilon_0^{ab}\right)\right\}\;,\] 
and
\[\left\{\otimes_i\cal F_{out}\left(\cal H_i\right)\right\}\bigotimes
\left\{\otimes_k\cal F_{out}\left(\cal H_k\right)\right\}\bigotimes
\cal F_{out}\left(\cal H\left(\varepsilon_0^{ab}\right)\right)\;,\] 
where $\cal H_i$ and $\cal H_k$ are the Hilbert spaces spanned by
$\left\{_i\lambda^a,{_i\tau^a}\right\}$ resp. $ \left\{_k\gamma^a\right\}$ and
$\cal H\left(\varepsilon_0^{ab}\right)$ is the Hilbert space spanned by all
other basis elements of $\cal H_{H^+}\oplus\cal H_{J^+}$, $\psi$ is mapped to
the following \underline{product} state
\[\psi\left(\varepsilon^{ab}\right)=\left(\otimes_i\psi_i\right)\bigotimes
\left(\otimes_k{\left(\psi_0\right)}_k\right)\bigotimes
\psi\left(\varepsilon^{ab}_0\right)\;,\]
where
\begin{equation}
\psi_i=\left(1,0,\sqrt{{1\over2}}e^{-{{\pi\omega_i}\over\kappa}}\,2\,
_i{\lambda^{\left(a\right.}}_i\tau^{\left.b\right)},0,\sqrt{{{3.1}\over{4.2}}}
e^{-2{{\pi\omega_i}\over\kappa}}\,4\,
{_i\lambda^{\left(a\right.}}{_i\tau^{b}}
{_i\lambda^{c}}{_i\tau^{\left.d\right)}},0,...\right)\;.
\end{equation}\\\\
${\left(\psi_0\right)}_k$ is the vacuum state of
$\cal F_{out}\left(\cal H_k\right)$
and
\[\psi=\psi\left(\varepsilon^{ab}_0\right)=\left(1,0,\sqrt{{1\over2}}
\varepsilon^{ab}_0,0,\sqrt{{3.1\over4.2}}\varepsilon^{\left(ab\right.}_0  
\varepsilon^{\left.cd\right)}_0,0,...\right)\;.\]\\\\
All information about the measurements on a given mode $i$ is thus contained in
the pure state $\psi_i\otimes{\left(\psi_0\right)}_i$ and there are no
correlations between the different modes because the total state is the product
of the states for the different modes.\\\\
To obtain the density matrix describing the measurements on a given mode $i$ at
$J^+$, we must trace over the Fock space
$\cal F_{\sigma_i}\left(\cal H_{\sigma_i}\right)\otimes
\cal F_{\tau_i}\left(\cal H_{\tau_i}\right)$ where $\cal H_{\sigma_i}$ and
$\cal H_{\tau_i}$ are spanned by $\sigma_i$ resp. $\tau_i$.\\\\
First, we trace over $\cal F_{\tau_i}$:
\[\begin{array}{rcl}
\tilde D_{\psi_i}&=&\sum_{N,N',M}\left<\lambda_i^N,\tau_i^M|\psi_i\right>
\left<\psi_i|\lambda_i^{N'},\tau_i^M\right>\left|\lambda_i^N\right>
\left<\lambda_i^{N'}\right|\\\\
\ &=&\sum_{N=0}^{+\infty}
{\left|\left<\lambda_i^N,\tau_i^N|\psi_i\right>\right|}^2
\left|\lambda_i^N\right>\left<\lambda_i^N\right|\;.
\end{array}\]\\\\
But ${\left|\left<\lambda_i^N,\tau_i^N|\psi_i\right>\right|}^2$, which is
proportional to the probability $\tilde P_N$ for observing N "particles" in
state $\lambda_i$, is obtained by taking the squared norm of the vector in
$\psi_i$ which is proportional to $\left|\lambda_i^{N},\tau_i^N\right>$
\begin{eqnarray}
\tilde P_N&\propto&\frac{\left(2N-1\right)\left(2N-3\right)...1}
{2N\left(2N-2\right)...2}e^{-2N{{\pi\omega_i}\over\kappa}}\,2^{2N}
{\left|\left|{_i\lambda^{\left(a_1\right.}}{_i\tau^{a_2}}...
{_i\lambda^{a_{2N-1}}}{_i\tau^{\left.a_{2N}\right)}}\right|\right|}^2\nonumber\\
\ &=&\frac{\left(2N-1\right)\left(2N-3\right)...1}
{2N\left(2N-2\right)...2}e^{-2N{{\pi\omega_i}\over\kappa}}\,2^{2N}
\frac{N!N!}{\left(2N\right)!}\nonumber\\
\ &=&e^{-N{{2\pi\omega_i}\over\kappa}}\;.
\end{eqnarray}\\\\
This is precisely the Boltzman factor with a temperature $T$ given
by
\begin{equation}
kT=\frac{\hbar\kappa}{2\pi}\;.
\end{equation}\\\\
The density matrix obtained by tracing over $_i\tau^a$ is therefore 
\[\tilde D_{\psi_i}=\sum_{n=0}^{+\infty}e^{-n{{\hbar\omega_i}\over kT}}
\left|\lambda_i^n\right>\left<\lambda_i^n\right|\;.\]\\\\
For each state $\left|\lambda_i^n\right>$,
we obtain a density matrix obtained by
tracing over $\cal F_{\sigma_i}$:
\[\begin{array}{rcl}
D_{\lambda_i^n}&=&\sum_{N,N',M}
\left<\rho_i^N,\sigma_i^M|\lambda_i^n\right>
\left<\lambda_i^n|\rho_i^{N'},\sigma_i^M\right>\left|\rho_i^N\right>
\left<\rho_i^{N'}\right|\\\\
\ &=&\sum_{N=0}^{n}
{\left|\left<\rho_i^N,\sigma_i^{n-N}|\lambda_i^n\right>\right|}^2
\left|\rho_i^N\right>\left<\rho_i^N\right|\;,
\end{array}\]
since $\lambda_i^n={\left(t_i\rho_i+r_i\sigma_i\right)}^n=
\sum_{N=0}^{n} \left(\!\!\begin{array}{c}n\\N\end{array}\!\!\right) 
t_i^Nr_i^{n-N}\rho_i^{\left(N\right.}\sigma_i^{\left.n-N\right)}$.\\\\
Setting
\[\Gamma_i={\left|t_i\right|}^2\Rightarrow1-\Gamma_i={\left|r_i\right|}^2\;,\] 
and using
\[{\left|\left|\left(\!\!\begin{array}{c}n\\N\end{array}\!\!\right)
\rho_i^{\left(N\right.}\sigma_i^{\left.n-N\right)}\right|\right|}^2=
\left(\!\!\begin{array}{c}n\\N\end{array}\!\!\right)\;,\] 
we obtain
\[D_{\lambda_i^n}=
\sum_{N=0}^{n}\left(\!\!\begin{array}{c}n\\N\end{array}\!\!\right)
\Gamma_i^N{\left(1-\Gamma_i\right)}^{n-N}
\left|\rho_i^N\right>\left<\rho_i^N\right|\;.\]\\\\
Defining $x=e^{-{{\hbar\omega_i}\over kT}}$, the density matrix describing the
measurements on a given mode $i$ at $J^+$ becomes
\[\begin{array}{rcl}
D_{\psi_i}&=&\sum_{n=0}^{+\infty}x^n D_{\lambda_i^n}=\sum_{n=0}^{+\infty}x^n
\sum_{N=0}^{n}\left(\!\!\begin{array}{c}n\\N\end{array}\!\!\right)
\Gamma_i^N{\left(1-\Gamma_i\right)}^{n-N}
\left|\rho_i^N\right>\left<\rho_i^N\right|\\\\
\ &=&\sum_{N=0}^{+\infty}\left|\rho_i^N\right>\left<\rho_i^N\right|
\sum_{n=N}^{+\infty}\left(\!\!\begin{array}{c}n\\N\end{array}\!\!\right)
\Gamma_i^N{\left(1-\Gamma_i\right)}^{n-N}x^n\\\\
\ &=&\sum_{N=0}^{+\infty}\left|\rho_i^N\right>\left<\rho_i^N\right|
\left\{{\left(\Gamma_ix\right)}^N\sum_{n'=0}^{+\infty}
\left(\!\!\begin{array}{c}{n'+N}\\N\end{array}\!\!\right)
{\left(\left(1-\Gamma_i\right)x\right)}^{n'}\right\}\\\\
\ &=&\sum_{N=0}^{+\infty}\left|\rho_i^N\right>\left<\rho_i^N\right|
\frac{{\left(\Gamma_ix\right)}^N}
{{\left(1-\left(1-\Gamma_i\right)x\right)}^{N+1}}\;.
\end{array}\]\\\\
Multiplying $D_{\psi_i}$ by $\left(1-\left(1-\Gamma_i\right)x\right)$, we
finally get
\begin{equation}
D_{\psi_i}=\sum_{N=0}^{+\infty}
{\left(\frac{\Gamma_i e^{-{{\hbar\omega_i}\over kT}}}
{1-\left(1-\Gamma_i\right)e^{-{{\hbar\omega_i}\over kT}}}\right)}^N
\left|\rho_i^N\right>\left<\rho_i^N\right|\;.
\end{equation}\\\\
Writing
\[x'=\frac{\Gamma_i e^{-{{\hbar\omega_i}\over kT}}}
{1-\left(1-\Gamma_i\right)e^{-{{\hbar\omega_i}\over kT}}}\;,\] 
we see that the probability for observing $N$ particles in the mode $i$
at $J^+$ is
\[P_N=\frac{x'^N}{\sum_{M=0}^{+\infty}x'^M}=\left(1-x'\right)x'^N\;,\]
and the average number of particles is
\begin{eqnarray}
\sum_{N=0}^{+\infty}N\left(1-x'\right)x'^N&=&x'\frac{\partial}{\partial x'}
\ln{1\over{1-x'}}=\frac{x'}{1-x'}\nonumber\\
\ &=&\Gamma_i\frac{e^{-{{\hbar\omega_i}\over kT}}}
{1-e^{-{{\hbar\omega_i}\over kT}}}\;.
\end{eqnarray}\\\\
The most important result of this section is that the black hole emits
particles with a {\em thermal spectrum} at temperature $kT=\hbar\kappa/2\pi$.
As a consequence, the quantum state describing these particles is mixed, that
is, it is described by a density matrix. If the black hole evaporates
completely, and Hawking's semiclassical calculation is assumed to be valid
throughout the evaporation process, then, starting from the pure state 
$\left|0_{in}\right>$ at $J^-$, it will evolve to the mixed thermal state at 
$J^+$ \cite{Hawk76}. The relation between incoming and outgoing state is
therefore not given
by a unitary $S$-matrix, and quantum mechanics appear to be violated: this is
the essence of Hawking's paradox.

\newpage
\section{Two Dimensional Black Holes}
\subsection{Classical Two Dimensional Black Holes} 
The main problem in the study of black hole evaporation in four 
dimensions is to include the back reaction of the matter fields on
the background geometry of the black hole: one would like to solve 
the semiclassical Einstein equations
\begin{equation}
G_{ab}=8\pi\left<\psi\right|\hat T_{ab}\left|\psi\right>\;,
\end{equation}
where $\left<\psi\right|\hat T_{ab}\left|\psi\right>$ is the expectation value
of the stress-energy tensor $\hat T_{ab}$ of the matter fields in a state
$\left|\psi\right>$.\\\\
$\left<\psi\right|\hat T_{ab}\left|\psi\right>$ is
divergent but it can be regularized. The divergent terms, which do not depend on
$\left|\psi\right>$, can be obtained from an effective action containing two
terms that are quadratic in the curvature tensor, see \cite[chapter 6]{Birr}.\\
Thus, the renormalized
expectation value $\left<\psi\right|\hat T_{ab}\left|\psi\right>_{ren}$ has a two
parameter ambiguity. These two terms are fourth order (i.e., they contain four
derivatives) and lead to instabilities in the semiclassical equations. Besides
that, $\left<\psi\right|\hat T_{ab}\left|\psi\right>_{ren}$ contains terms which
are nonlocal so that in four dimensions, the problem  seems intractable.\\\\
On the other hand, in two dimensions, the renormalized expectation value of the
stress-energy tensor contains two derivatives of the metric, so the
semiclassical Einstein equations remain second order.\\\\
In two dimensions, the Einstein-Hilbert action
\[S={1\over4\pi}\int d^2x\,\sqrt{-g}R\;,\]
is a topological invariant because
\[\delta S={1\over4\pi}\int d^2x\,\sqrt{-g}\left\{\left(R_{ab}-{1\over2}g_{ab}R
\right)\delta g^{ab}+g_{ab}\nabla^2\delta g^{ab}-\nabla_a\nabla_b\delta
g^{ab}\right\}\;,\]
where we have used
\[\left\{\begin{array}{rcl}
g^{ab}\delta R_{ab}&=&g_{ab}\nabla^2\delta g^{ab}-\nabla_a\nabla_b\delta g^{ab}
\\
\delta \sqrt{-g}&=&-{1\over2}\,\sqrt{-g}\,g_{ab}\,\delta g^{ab}\;.
\end{array}\right.\]
In two dimensions
\[R_{ab}\equiv{1\over2}\,g_{ab}\,R\;,\]
for all metrics. Therefore, $\delta S$ is the integral of a total divergence and
we have no classical equations of motion. It can be shown that for a compact two
dimensional manifold of genus $g$, the Einstein action gives the so-called Euler
characteristic
\[{1\over4\pi}\int d^2x\,\sqrt{g}R=\chi=2\left(1-g\right)\;.\]\\
To get a dynamical theory of gravity in two dimensions, one can couple gravity
to the scalar dilaton field $\Phi$. This coupling is realized in string theory, 
where $\Phi$ is one of the massless modes of the closed string together with the
graviton and an antisymmetric tensor field. In the low energy limit, the
Lagrangian describing this interaction is, in four dimensions,
\cite[chapter 3]{GSW}
\[S=\int d^4x\,\sqrt{-g}e^{-2\Phi}\left\{R+4{\left(\nabla\Phi\right)}^2-
{1\over2}g^{ae}g^{bf}F_{ab}F_{ef}\right\}\;,\]
where we have added a Maxwell field associated with a $U(1)$ subgroup of 
$E_8\otimes E_8$ or $spin\left(32\right)/Z_2$ and we have set to zero the
remaining gauge fields and antisymmetric tensor field.\\\\
Defining a new metric
$\tilde g$, with $\tilde g=e^{-2\Phi}g$, the action becomes
\[S=\int d^4x\,\sqrt{-\tilde g}\left\{\tilde R-2\tilde g^{ab}
\nabla_a\Phi\nabla_b\Phi-
{1\over2}e^{-2\Phi}\tilde g^{ae}\tilde g^{bf}F_{ab}F_{ef}\right\}\;.\]
Here we have used that
\[\left\{\begin{array}{lll}
\sqrt{-g}&=&e^{4\Phi}\sqrt{-\tilde{g}}\\\\
R&=&e^{-2\Phi}\left(\tilde{R}-6\,\tilde{g}^{ab}\tilde{\nabla}_{a}
\tilde{\nabla}_{b}\Phi-6\,\tilde{g}^{ab}\tilde{\nabla}_{a}\Phi
\tilde{\nabla}_{b}\Phi\right)\;,
\end{array}\right.\]
if $g=e^{+2\Phi}\tilde g$\, in four dimensions.\\\\
When $F_{ab}=0$, this action describes ordinary Einstein gravity coupled to
a massless Klein-Gordon field. The no-hair theorems then imply that the unique
stationary black hole solutions are described by the Kerr metric
~(\ref{eq:Kerr}) with zero charge.\\\\
When $F_{ab}\neq0$, $\Phi\neq0$,
and the solution is not of the Kerr form: the metric, dilaton and 
electromagnetic
field tensor for a non-rotating, magnetically charged and spherically symmetric
black hole are \cite{GHS}:
\[\left\{\begin{array}{lcl}
d\tilde s^2&=&\tilde g_{ab}dx^adx^b\\
\ &=&-\left(1-{2M\over r}\right)dt^2+
{\left(1-{2M\over r}\right)}^{-1}dr^2+r\left(r-{Q^2\over{2M}}e^{-2\Phi_0}\right)
d\Omega^2\nonumber\\\\
e^{-2\Phi}&=&e^{-2\Phi_0}\left(1-{Q^2\over{2Mr}}e^{-2\Phi_0}\right)\\\\
F_{ab}&=&Q\sin{\theta}\,2\,d\theta_{\left[a\right.}d\phi_{\left.b\right]}\;.
\end{array}\right.\]
Here, $\Phi_0$ is the value of the dilaton at infinity.\\\\
This metric is almost the same as the Schwarzschild metric. The only difference
is that the areas of the two-spheres are decreased relative to the
Schwarzschild values
\[{A\over4\pi}=r^2\left(1-{Q^2\over2Mr}e^{-2\Phi_0}\right)=
r^2e^{-2\left(\Phi-\Phi_0\right)}\;.\]\\
The area is zero when $r={Q^2\over{2M}}e^{-2\Phi_0}$, causing this surface to
be singular, but the horizon is still located at $r = 2M$.\\\\
When 
${Q^2\over{2M}}e^{-2\Phi_0}<2M$, the singularity is inside the horizon; when
${Q^2\over{2M}}e^{-2\Phi_0}=2M$, the horizon becomes singular and when 
${Q^2\over{2M}}e^{-2\Phi_0}>2M$ the solution has a naked singularity.\\\\
Comparing with the charged solutions of the Einstein-Maxwell theory, given by
the Reissner-Nordstrom metric (\ref{eq:ReiNo}), we note several differences:\\
First, when the dilaton is present, there is no inner horizon.\\
Second, the
transition from black hole to naked singularity occurs at
${Q^2\over{2M}}e^{-2\Phi_0}=2M$ rather than $Q^2=M^2$ as for Reissner-Nordstrom
black holes and furthermore, at this transition (the black hole
is then called an extremal hole), the horizon is singular in the
first case, but completely regular in the second case.\\\\ 
However, for extremal dilaton black holes, the metric seen by the string is 
$g=e^{2\Phi}\tilde g$, i.e., 
\[ds^2=g_{ab}dx^adx^b=e^{2\Phi_0}\left(-dt^2+
{\left(1-{{2M}\over r}\right)}^{-2}dr^2+r^2d\Omega^2\right)\;.\]\\
Defining the coordinates $\tau$, $\sigma$ by
\[\left\{\begin{array}{lclcl}
\tau&=&e^{\Phi_0}\,t&\ &\\\\
d\sigma&=&e^{\Phi_0}{\left(1-{{2M}\over r}\right)}^{-1}dr&\Rightarrow&
\sigma=e^{\Phi_0}\left(r+2M\ln\left({r\over2M}-1\right)\right)\;,
\end{array}\right.\]
we get
\[\left\{\begin{array}{lcccl}
ds^2&=&-d\tau^2+d\sigma^2+e^{2\Phi_0}\,r^2\,d\Omega^2&\ &\\\\
e^{-2\Phi}&=&e^{-2\Phi_0}\left(1-{2M\over r}\right)&=&e^{-2\Phi_0}\,{2M\over r}
\,\exp{\left(\frac{e^{-\Phi_0}\,\sigma-r}{2M}\right)}\;.
\end{array}\right.\]\\\\
When $r\rightarrow+\infty$, the geometry approaches that of flat four
dimensional space;\\
when $r\rightarrow 2M$ or $\sigma\rightarrow-\infty$, it
approaches that of flat two dimensional space in the variable
$\left(\sigma,\tau\right)$ times a two sphere of radius 
$R=2Me^{\Phi_0}=\left|Q\right|$ and
\[-2\Phi\sim\frac{\sigma}{2M\,e^{\Phi_0}}\Leftrightarrow
\Phi\sim-\lambda\sigma\;,\]
where $\lambda={1\over2Q}$.\\\\
Thus, the geometry seen by the string is free of horizons and singularities in
contrast to extremal Reissner-Nordstrom black holes.
This geometry is that of a bottomless hole, see fig.~\ref{fig:hole}.
\begin{figure}
	\vspace{6 cm}
	\includegraphics{figbh/Hole.eps}
	\caption{}
	\label{fig:hole}
\end{figure}
In the throat region, where $r\rightarrow 2M$, the dilaton field is linear in
$\sigma$. The extremal dilaton black hole is consequently called the linear
dilaton vacuum. It is called vacuum because it is stable, i.e., it does not emit
any Hawking radiation. This is because the string metric is free of horizons and
singularities and thus, the Hawking temperature is zero, as for extremal
Reissner-Nordstrom black holes (see~(\ref{eq:kappa})).\\\\
For non-extremal black holes,
the string metric still describes a black hole with an event horizon and a
singularity because the conformal factor $e^{2\Phi}$ is finite everywhere 
outside (and on) the horizon.\\\\
At low energies compared to
${1\over R}$, the physics in the throat region of an extremal or
near extremal black hole can be described by an
effective two dimensional action
\[S={1\over2\pi}\int d^2x\,\sqrt{-g}e^{-2\Phi}\left\{R+
4{\left(\nabla\Phi\right)}^2+4\lambda^2-{1\over2}F^2\right\}\;.\]
Because the electromagnetic field tensor $F$ has no propagating degrees of
freedom in two dimensions, we can set it to zero if no charged particles are
present.\\\\
We will see later that this two dimensional model indeed has a linear
dilaton vacuum solution and also that a particle thrown into this dilaton
vacuum turns it into a non-extremal black hole with a horizon and a
singularity.
\\\\
We note that in the throat region of a near extremal black hole, the area of
two-spheres, as measured in the canonical metric $\tilde g$, is
\[{A\over4\pi}=r^2e^{-2\left(\Phi-\Phi_0\right)}
\approx4M^2e^{2\Phi_0}e^{-2\Phi}\approx Q^2e^{-2\Phi}\;.\]\\
The area of two-spheres in the four dimensional theory is thus proportional
to $e^{-2\Phi}$. This fact will be useful later when we will study the two
dimensional theory without refering to its four dimensional origin.\\\\
Thus, we
will start from the action
\begin{equation}
{1\over2\pi}\int d^2x\,\sqrt{-g}\,e^{-2\Phi}\left\{R+
4{\left(\nabla\Phi\right)}^2+4\lambda^2\right\}\;,
\end{equation}
and look at it as a model of quantum gravity in two dimensions. This model was
proposed in 1991 by Callan, Giddings, Harvey and Strominger
\cite{CGHS} and is therefore
usually referred to as the CGHS model.\\\\
Another model of
two dimensional gravity is the one obtained from the four dimensional
Einstein-Hilbert action by restricting oneself to spherical symmetric metrics
(For the definition of $x^{\pm}$, see (\ref{eq:null}))
\[\begin{array}{lclcl}
S_4&\propto&\int d^4x\sqrt{-g_4}\,R_4\;,&\ &\\\\
ds^2&=&{\left(g_4\right)}_{ab}\,dx^adx^b&=&-e^{2\rho}dx^+dx^-+
e^{-2\Phi}d\Omega^2\;,
\end{array}\]
or
\[\left({\left(g_4\right)}_{ab}\right)=
\left(\begin{array}{cccc}
0&-{1\over2}e^{2\rho}&0&0\\
-{1\over2}e^{2\rho}&0&0&0\\
0&0&e^{-2\Phi}&0\\
0&0&0&e^{-2\Phi}\sin^2\theta
\end{array}\right)\;.\] 
This implies
\[\sqrt{-g_4}=\sqrt{-g_2}\,e^{-2\Phi}\sin\theta\;.\]
It can also be shown that (see \cite[exercise 14.16]{grav})
\[\begin{array}{lcl}
R_4&=&2e^{2\Phi}+e^{-2\rho}\left(8\partial_+\partial_-\rho+
24\partial_+\Phi\partial_-\Phi-16\partial_+\partial_-\Phi\right)\\\\
\ &=&2e^{2\Phi}+R_2-6{\left(\nabla\Phi\right)}^2+4\nabla^2\Phi\;,
\end{array}\]
where we have used that $R_2=8\,e^{-2\rho}\partial_+\partial_-\rho$
(see~(\ref{eq:curv})).\\\\
Therefore
\[S_2\propto\int d^2x\,\sqrt{-g_2}\,e^{-2\Phi}\left\{R_2+
2{\left(\nabla\Phi\right)}^2+2e^{2\Phi}\right\}\;,\]
where we have integrated by parts once.
This action looks similar to the CGHS action with the same interpretation 
for $e^{-2\Phi}$.\\\\
The CGHS action is
\begin{equation}
\label{eq:CGHSa}
S={1\over2\pi}\int d^2x\,\sqrt{-g}\left\{e^{-2\Phi}\left(R+
4{\left(\nabla\Phi\right)}^2+4\lambda^2\right)
-{1\over2}\sum_{i=1}^{N}{\left(\nabla f_i\right)}^2\right\}\;,
\end{equation}
where we have added $N$ scalar massless matter fields.\\\\
To obtain the classical
equations of motion for the metric $g$, we vary $S$ with respect to $g$. We use
the following relations
\[\begin{array}{ccccc}
\delta \sqrt{-g}&=&-{1\over2}\sqrt{-g}\,g_{ab}\,\delta g^{ab}&\ &\\
\delta R&=&\delta \left(R_{ab}\,g^{ab}\right)&=&\delta R_{ab}\,g^{ab}+R_{ab}\,
\delta g^{ab}\;,\end{array}\]
where
\[g^{ab}\delta R_{ab}=g_{ab}{\nabla}^2\delta g^{ab}-\nabla_a\nabla_b\delta
g^{ab}\;,\]
and
\[R_{ab}={1\over2}g_{ab}R\;.\]\\
Then
\[\begin{array}{lcl}
\delta_gS&=&{1\over2\pi}\int d^2x\sqrt{-g}\left\{
-{1\over2}\left(e^{-2\Phi}\left(R+4{\left(\nabla\Phi\right)}^2+
4\lambda^2\right)-{1\over2}{\left(\nabla f_i\right)}^2\right)g_{ab}\right.\\\\
\ &+&\left.\left(e^{-2\Phi}\,4\,\nabla_a\Phi\nabla_b\Phi-
{1\over2}\,\nabla_af_i\nabla_bf_i\right)\right.\\\\
&+&\left.\left(e^{-2\Phi}{1\over2}\,g_{ab}R+g_{ab}{\nabla}^2e^{-2\Phi}-
\nabla_a\nabla_be^{-2\Phi}\right)\right\}\delta g^{ab}\\\\
&=&{1\over2\pi}\int d^2x\sqrt{-g}\left\{-{1\over2}\left(e^{-2\Phi}
\left(R+4{\left(\nabla\Phi\right)}^2+4\lambda^2\right)-
{1\over2}{\left(\nabla f_i\right)}^2\right)g_{ab}\right.\\\\
&+&\left(e^{-2\Phi}\,4\,\nabla_a\Phi\nabla_b\Phi
-{1\over2}\,\nabla_af_i\nabla_bf_i\right)\\\\
\ &+&\left.e^{-2\Phi}\left({1\over2}\,g_{ab}R+
\left(4\,{\left(\nabla\Phi\right)}^2-
2\,{\nabla}^2\Phi\right)g_{ab}-\left(4\nabla_a\Phi\nabla_b\Phi-2\nabla_a\nabla_b
\Phi\right)\right)\right\}\delta g^{ab}\;.
\end{array}\]\\
Setting the variation to zero, we obtain the equations of motion
\[\begin{array}{cl}
\ &e^{-2\Phi}\left(2\nabla_a \nabla_b\Phi-2g_{ab}\left(\nabla^2\Phi-
{\left(\nabla\Phi\right)}^2+\lambda^2\right)\right)\\\\
=&{1\over2}\left(\nabla_a f_i\nabla_b f_i-{1\over2}{\left(\nabla f_i\right)}^2
g_{ab}\right)\equiv T^f_{ab}\;.\end{array}\]\\
These equations correspond to the Einstein equations in ordinary four
dimensional gravity.\\\\
Varying the action with respect to $\Phi$
\[\begin{array}{lcl}
\delta_{\Phi}S&=&{1\over2\pi}\int d^2x\sqrt{-g}\left\{\left(R+
4\,{\left(\nabla\Phi\right)}^2+4\lambda^2\right)\delta e^{-2\Phi}+e^{-2\Phi}
4\,\delta{\left(\nabla \Phi\right)}^2\right\}\\\\
&=&{1\over2\pi}\int d^2x\sqrt{-g}\left\{-2e^{-2\Phi}\left(R+
4{\left(\nabla\Phi\right)}^2+4\lambda^2\right)-8\nabla.\left(e^{-2\Phi}
\nabla\Phi\right)\right\}\delta\Phi\\\\
&=&{1\over2\pi}\int d^2x\sqrt{-g}e^{-2\Phi}
\left\{-2\left(R+4{\left(\nabla\Phi\right)}^2+4\lambda^2\right)\right.\\\\
&-&8\left.\left(\nabla ^2\Phi-2{\left(\nabla\Phi\right)}^2\right)\right\}
\delta\Phi=0\;,\end{array}\]
one gets
\[{R\over4}+\nabla^2\Phi-{\left(\nabla\Phi\right)}^2+\lambda^2=0\;.\]\\
Finally
\[\delta_f S={1\over2\pi}\int d^2x\sqrt{-g}\,\nabla^2f_i\,\delta f_i=0\;,\]
or
\[\nabla^2f_i=0\;.\]\\
The equations of motion are thus
\begin{equation}
\left\{\begin{array}{ccc}
2e^{-2\Phi}\left(\nabla_a \nabla_b\Phi-g_{ab}\left(\nabla^2\Phi-
{\left(\nabla\Phi\right)}^2+\lambda^2\right)\right)&=&T_{ab}^f\nonumber\\\\
{R\over4}+\nabla^2\Phi-{\left(\nabla\Phi\right)}^2+\lambda^2&=&0\nonumber\\\\
\nabla^2f_i&=&0\;.
\end{array}\right.
\end{equation}\\\\
Since a reparametrization depends on two arbitrary functions of the old 
coordinates we can always find a coordinate system in which the two diagonal
elements of $g$ are vanishing. The new coordinates are therefore null
coordinates and we call them $x^{\pm}$. Thus
\begin{eqnarray}
\label{eq:null}
g_{++}&=&g_{--}=0\;,\nonumber\\
g_{+-}&=&-{1\over2}\,e^{2\rho}\;,
\end{eqnarray}
or
\[ds^2=-e^{2\rho}\,dx^+dx^-\;.\]
This is called the conformal gauge.\\\\
In this gauge, the non-vanishing connections
are
\[\Gamma^+_{++}=2\,\partial_+\rho,\ \Gamma^-_{--}=2\,\partial_-\rho\;,\]
which gives
\[\left[\nabla_+,\nabla_-\right]\omega_+=\left[\partial_+-2\partial_+\rho,
\partial_-\right]\omega_+=2\,\partial_+\partial_-\rho\,\omega_+\;.\]\\\\
By definition
\[\left[\nabla_+,\nabla_-\right]\omega_+=R_{+-+}^{\ \ \ \ \ +}\omega_+\;,\]
that is
\[R_{+-+}^{\ \ \ \ \ +}=2\,\partial_+\partial_-\rho\;,\]
and
\[\begin{array}{lcl}
R_+^{\ \ +}&=&R_{++}^{\ \ \ \ ++}+R_{+-}^{\ \ \ \ +-}
=-g^{-+}R_{+-+}^{\ \ \ \ \ +}=4e^{-2\rho}\partial_+\partial_-\rho\;,\\\\
R_-^{\ \ -}&=&R_+^{\ \ +}\;.
\end{array}\]\\\\
Therefore, the curvature scalar $R$ is
\begin{equation}
\label{eq:curv}
R=R_+^{\ \ +}+R_-^{\ \ -}=8\,e^{-2\rho}\partial_+\partial_-\rho\;.
\end{equation}\\\\
In the conformal gauge, the equations of motion are:
\[\begin{array}{lcl}
g_{+-}&:&2e^{-2\Phi}\left(\partial_+\partial_-\Phi
-g_{+-}\left(2g^{+-}\partial_+\partial_-\Phi-2g^{+-}\partial_+
\Phi\partial_-\Phi+\lambda^2\right)\right)\\\\
\ &=&{1\over2}\left(\partial_+f_i\partial_-f_i-{1\over2}g_{+-}\,2g^{+-}
\partial_+f_i\partial_-f_i\right)\\\\
\ &\Leftrightarrow&4\partial_+\Phi\partial_-\Phi-2\partial_+\partial_-\Phi+
\lambda^2e^{2\rho}=0\\\\
\Phi&:&2e^{-2\rho}\partial_+\partial_-\rho-4e^{-2\rho}\partial_+\partial_-\Phi
+4e^{-2\rho}\partial_+\Phi\partial_-\Phi+\lambda^2=0\\\\
\ &\Leftrightarrow&2\partial_+\partial_-\rho-4\partial_+\partial_-\Phi+
4\partial_+\Phi\partial_-\Phi+\lambda^2e^{2\rho}=0\\\\
f_i&:&\partial_+\partial_-f_i=0\;.
\end{array}\]\\\\
The remaining equations for $g_{++}$ and $g_{--}$ are constraints coming from
the gauge fixing $g_{++}=g_{--}=0$
\[\begin{array}{lcl}
g_{\pm\pm}&:&2e^{-2\Phi}\nabla_{\pm}\nabla_{\pm}\Phi=
{1\over2}\partial_{\pm}f_i\partial_{\pm}f_i\\\\
\ &\Leftrightarrow&e^{-2\Phi}\left(2\partial_{\pm}^2\Phi-4\partial_{\pm}\rho
\partial_{\pm}\Phi\right)={1\over2}\partial_{\pm}f_i\partial_{\pm}f_i\;.
\end{array}\]\\\\
So
\begin{equation}
\left\{\begin{array}{cccc}
4\partial_+\Phi\partial_-\Phi-2\partial_+\partial_-\Phi+
\lambda^2e^{2\rho}&=&0&\left(\rho\right)\\\\
2\partial_+\partial_-\rho-4\partial_+\partial_-\Phi+
4\partial_+\Phi\partial_-\Phi+\lambda^2e^{2\rho}&=&0&\left(\Phi\right)\\\\
\partial_+\partial_-f_i&=&0&\left(f_i\right)\\\\
e^{-2\Phi}\left(2\partial_{\pm}^2\Phi-4\partial_{\pm}\rho
\partial_{\pm}\Phi\right)&=&{1\over2}\partial_{\pm}f_i\partial_{\pm}f_i
&\left(g_{\pm\pm}\right)\;.
\end{array}\right.
\end{equation}\\\\
Substracting $\left(\rho\right)$ from $\left(\Phi\right)$ we get
\[\Leftrightarrow\begin{array}{ccl}
\partial_+\partial_-\left(\rho-\Phi\right)&=&0\\
\rho-\Phi&=&f_+\left(x^+\right)+f_-\left(x^-\right)\;.
\end{array}\]\\\\
The conformal gauge is preserved by arbitrary coordinate transformations of the
form
\[\left\{\begin{array}{l}
x^+\rightarrow\tilde x^+\left(x^+\right)\\
x^-\rightarrow\tilde x^-\left(x^-\right)\;,
\end{array}\right.\]
since
\[0=g_{\pm\pm}\left(x\right)={\left(\frac{d\tilde x^{\pm}}{dx^{\pm}}\right)}^2
\tilde g_{\pm\pm}\left(\tilde x\right)\;.\]\\\\
This fact allows us to choose coordinates such that $f_+\left(x^+\right)=
f_-\left(x^-\right)=0$, so we can set $\Phi=\rho$. For reasons that will be
clear in the following, these coordinates will be called the
Kruskal coordinates.\\\\
The equations of motion and constraints become:
\begin{equation}
\label{ClCGHS}
\left\{\begin{array}{cccc}
-\partial_+\partial_-e^{-2\rho}&=&\lambda^2&\hspace{1cm}\left(\rho\right)\\\\
\partial_+\partial_-f_i&=&0&\hspace{1cm}\left(f_i\right)\\\\
-\partial_{\pm}^2e^{-2\rho}&=&T_{\pm\pm}^f\left(={1\over2}\partial_{\pm}f_i
\partial_{\pm}f_i\right)&\hspace{1cm}\left(g_{\pm\pm}\right)\;.
\end{array}\right.
\end{equation}\\\\
The general static solution when $f_i=0$ is
\[e^{-2\Phi}=e^{-2\rho}={M\over\lambda}-\lambda^2x^+x^-\;.\]
In the following, $M$ is no longer the mass of the four dimensional black hole
although we use the same letter. In fact, this new $M$ turns out to be the mass
of the {\em two dimensional} black hole.\\\\
When $M=0$
\[ds^2=-e^{2\rho}dx^+dx^-=-{\left(-\lambda^2x^+x^-\right)}^{-1}dx^+dx^-\;.\]\\
Defining the coordinates $\sigma^{\pm}$ by
\[\lambda x^{\pm}=\pm e^{\pm\lambda\sigma^{\pm}}\Rightarrow dx^{\pm}=
\pm\lambda x^{\pm}d\sigma^{\pm}\;,\]
we see that the spacetime is flat:
\[ds^2=-d\sigma^+d\sigma^-=-d\tau^2+d\sigma^2\;,\]
where
\[\sigma^{\pm}=\tau\pm\sigma\;.\]\\
Furthermore
\[e^{-2\Phi}=-\lambda^2x^+x^-=e^{\lambda\left(\sigma^+-\sigma^-\right)}=
e^{2\lambda\sigma}\;,\]
or
\[\Phi=-\lambda\sigma\;.\]\\
We thus recognize the linear dilaton vacuum described earlier. In the
four dimensional theory, this solution corresponds to an extremal dilaton
black hole solution.\\\\
When $M\neq0$, the scalar curvature is
\begin{eqnarray}
R&=&8e^{-2\rho}\partial_+\partial_-\rho=4\left(e^{2\rho}\partial_+e^{-2\rho}
\partial_-e^{-2\rho}-\partial_+\partial_-e^{-2\rho}\right)\nonumber\\
&=&4\left({\left({M\over\lambda}-\lambda^2x^+x^-\right)}^{-1}
\left(-\lambda^2x^-\right)\left(-\lambda^2x^+\right)
+\lambda^2\right)\nonumber\\
&=&\frac{4M\lambda}{{M\over\lambda}-\lambda^2x^+x^-}\;.
\end{eqnarray}\\
At $\lambda^3x^+x^-=M$, we have a curvature singularity asymptotically
aproaching the null curves $x^{\pm}=0$, which therefore are event horizons. The
spacetime diagram of this black hole solution, depicted in
fig.~\ref{fig:2DSch}, is qualitatively similar to that 
of the extended Schwarzschild solution reported in fig.~\ref{fig:Sch1}.
\begin{figure}
	\vspace{6 cm}
	\includegraphics{figbh/2DSch.eps}
	\caption{}
	\label{fig:2DSch}
\end{figure}
\\\\In the four dimensional theory, these black hole solutions correspond to 
near extremal dilaton black holes.\\\\
Using the coordinates $\sigma^{\pm}$
introduced earlier, we see that
\[\begin{array}{ccccc}
ds^2&=&-{\left({M\over\lambda}-\lambda^2x^+x^-\right)}^{-1}dx^+dx^-&\ &\\\\
\ &=&-{\left({M\over\lambda}+
e^{\lambda\left(\sigma^+-\sigma^-\right)}\right)}^{-1}
e^{\lambda\left(\sigma^+-\sigma^-\right)}d\sigma^+d\sigma^-&\ &\\\\
\ &=&-{\left({M\over\lambda}e^{-2\lambda\sigma}+1\right)}^{-1}d\sigma^+d\sigma^-
&=&-e^{2\rho'}d\sigma^+d\sigma^-\;.
\end{array}\]\\
Since the components of the metric do not depend on $\tau$,
${\left(\frac{\partial}{\partial{\tau}}\right)}^a$ is the timelike Killing
field of the manifold. Therefore, $\tau$ corresponds to $t$ in the Schwarzschild
metric.\\\\
We also see that region $I$ of fig.~\ref{fig:2DSch} is flat at spatial
and null infinity $J^{\pm}_R$: as $\sigma\rightarrow+\infty$
\[2\rho'\sim-{M\over\lambda}e^{-2\lambda\sigma}\]
which vanishes exponentially.\\\\
$\sigma$ corresponds to $r^*$ in the Schwarzschild metric. Thus, $\sigma^{\pm}$
correspond to 
$\left\{\begin{array}{c}v\\u\end{array}\right\}$ and $x^{\pm}$ correspond to
$\left\{\begin{array}{c}V\\U\end{array}\right\}$.\\\\
Also, when $\sigma\rightarrow+\infty$
\[\begin{array}{ll}
\ &e^{-2\Phi}={M\over\lambda}-\lambda^2x^+x^-={M\over\lambda}+
e^{2\lambda\sigma}\\\\
\Rightarrow&e^{-2\left(\Phi+\lambda\sigma\right)}=1+{M\over\lambda}
e^{-2\lambda\sigma}\\\\
\Rightarrow&\Phi\sim-\lambda\sigma-{M\over{2\lambda}}e^{-2\lambda\sigma}\;.
\end{array}\]\\\\
Therefore, we have shown that region $I$ of the two dimensional black hole
asymptotically approaches the linear dilaton vacuum at null infinity.\\\\
We can patch together the vacuum solution with the black hole solution across
some light-like line $x^+=x^+_0$ as follows
\begin{equation}
\label{eq:2Dcoll}
e^{-2\Phi}=e^{-2\rho}=\left\{
\begin{array}{ccc}
-\lambda^2x^+x^-&,&x^+<x^+_0\\
{M\over\lambda}-\lambda^2x^+\left(x^-+{M\over{\lambda^3x^+_0}}\right)
&,&x^+>x^+_0\;.
\end{array}\right.
\end{equation}
\begin{figure}[t]
	\vspace{6 cm}
	\includegraphics{figbh/2Dcoll1.eps}
	\caption{}
	\label{fig:2Dcoll1}
\end{figure}
\begin{figure}[t]
	\vspace{6 cm}
	\includegraphics{figbh/2Dcoll2.eps}
	\caption{}
	\label{fig:2Dcoll2}
\end{figure}
\\The corresponding spacetime diagram and Penrose diagram are shown
in fig.~\ref{fig:2Dcoll1} resp. fig.~\ref{fig:2Dcoll2}.\\\\
For $x^+>x^+_0$, the solution is identical to the black hole solution but it is
translated by $\Delta x^-=-{M\over{\lambda^3x^+_0}}$. In particular, the event
horizon is located at $x^-=-{M\over{\lambda^3x^+_0}}$.\\\\
There is a discontinuity
in $\partial_+e^{-2\rho}$:
\[\begin{array}{cccc}
\ &\partial_+e^{-2\rho}&=&\left\{
\begin{array}{ccc}
-\lambda^2x^-&,&x^+<x^+_0\\
-\lambda^2\left(x^-+{M\over{\lambda^3x^+_0}}\right)&,&x^+>x^+_0
\end{array}\right.\\\\
\Rightarrow&-\partial_+^2e^{-2\rho}&=&{M\over{\lambda x_0^+}}\,
\delta\left(x^+-x_0^+\right)\;.
\end{array}\]\\
The constraint $\left(g_{++}\right)$ is thus satisfied for
\[T_{++}^f={1\over2}\partial_+f\partial_+f={M\over{\lambda x_0^+}}
\delta\left(x^+-x_0^+\right)\;,\]
where we have specialized to a single massless scalar field $f$.\\\\
The asymptotically minkowskian coordinates $\sigma^{\pm}$ are now given by
\[\left\{\begin{array}{ccc}
\lambda x^+&=&e^{\lambda\sigma^+}\\
\lambda\left(x^-+{M\over{\lambda^3x^+_0}}\right)&=&-e^{-\lambda\sigma^-}\;,
\end{array}\right.\]
and they cover region $I$ of fig.~\ref{fig:2Dcoll2}.\\\\
The energy carried by the incoming field $f$ is most easily defined and
calculated in
the asymptotically flat region $J^-_R$. We use the asymptotically minkowskian
coordinates $\left(\tau,\;\sigma\right)$ because energy is defined as the
component of energy-momentum along minkowskian time, that is, $\tau$.\\\\
The energy-momentum flow through the volume one-form
\[\Sigma_a=\varepsilon_{ab}{\left(\frac{\partial}{\partial\sigma^+}\right)}^b
=\varepsilon_{\sigma^-\sigma^+}{\left(d\sigma^-\right)}_a=\sqrt{-g}\,
{\left(d\sigma^-\right)}_a={1\over2}\,{\left(d\sigma^-\right)}_a\;,\]
is, by definition \cite[chapter 5]{grav}
\[T^{ab}\Sigma_b=T^{\sigma^-\sigma^-}
{\left(\frac{\partial}{\partial\sigma^-}\right)}^a
{\left(\frac{\partial}{\partial\sigma^-}\right)}^b\,{1\over2}\,
{\left(d\sigma^-\right)}_b={1\over2}\,T^{\sigma^-\sigma^-}
{\left(\frac{\partial}{\partial\sigma^-}\right)}^a\;.\]\\
Since $\sigma^{\pm}=\tau\pm\sigma$ or
${\left(\partial/\partial\sigma^{\pm}\right)}^a=
{1\over2}\left({\left(\partial/\partial\tau\right)}^a\pm
{\left(\partial/\partial\sigma\right)}^a\right)$, 
the energy flow in the $\sigma^-$-direction per unit length in the $\sigma^+$-
direction is
\[{1\over4}\,T^{\sigma^-\sigma^-}={\left(g_{\sigma^+\sigma^-}\right)}^2\,
T^{\sigma^-\sigma^-}=T_{\sigma^+\sigma^+}\;.\]\\\\
The total energy carried by $f$ is
\[\begin{array}{lcl}
\int_{-\infty}^{+\infty}d\sigma^+T_{\sigma^+\sigma^+}^f&=&
\int_{-\infty}^{+\infty}dx^+\left(\frac{d\sigma^+}{dx^+}\right)
{\left(\frac{dx^+}{d\sigma^+}\right)}^2T_{++}^f\\\\
&=&\int_{-\infty}^{+\infty}dx^+\lambda x^+{M\over{\lambda x^+_0}}
\delta\left(x^+-x^+_0\right)=M\;.
\end{array}\]\\
Hence, the energy carried by the massless field is $M$; by energy
conservation, it is also equal to the mass of the two dimensional black hole;
this can be verified by calculating the Arnowitt-Deser-Misner (ADM) mass
of the black hole as done in \cite{Wit}.

\newpage
\subsection{The Trace Anomaly and Hawking Radiation}
Classically, the trace of the matter stress-energy tensor is zero since
\[\Rightarrow\begin{array}{l}
T_{ab}^f={1\over2}\left(\nabla_a f\nabla_b f-{1\over2}
{\left(\nabla f\right)}^2g_{ab}\right)\\\\
T^{a}_{\ \ a}={1\over2}\left({\left(\nabla f\right)}^2-
{\left(\nabla f\right)}^2\right)=0\;.
\end{array}\]
(we suppress the label $f$ in the following).
\\\\This reflects the conformal invariance of the matter action:
under a change
\[g_{ab}\rightarrow e^{2\alpha}g_{ab}\;,\]
which, only in two dimensions, also implies
\[\sqrt{-g}\rightarrow e^{2\alpha}\sqrt{-g}\;,\]
the matter action
\[S_m=-{1\over2\pi}\int d^2x\sqrt{-g}\,g^{ab}\,{1\over2}\,\nabla_a f
\nabla_b f\;,\]
changes to
\[S'_m=-{1\over2\pi}\int d^2x\left(e^{2\alpha}\sqrt{-g}\right)
\left(e^{-2\alpha}g^{ab}\right){1\over2}\,\nabla_a f
\nabla_b f=S_m\;.\]\\
On the other hand, for an infinitesimal change
$\delta g_{ab}=2\alpha\,g_{ab}$\,, the change in $S_m$ is
\[\delta S_m=-{1\over2\pi}\int d^2x\sqrt{-g}\,T_{ab}\,\delta g^{ab}=
{1\over2\pi}\int d^2x\sqrt{-g}\,T^a_{\ \ a}\,2\alpha\;,\]
which vanishes only if $T^a_{\ \ a}\equiv0$.\\\\
In curved spacetime, however, the conformal invariance is broken by the quantum
fluctuations of the matter field(s): the expectation value of
the trace of the stress-energy tensor turns out to be proportional to
the curvature scalar.\\\\
We now show this statement in the limit of a weak
curvature (we will follow the derivation of \cite[pages 185-188]{Cardy},
see also \cite[pages 468-472]{Alv}).
For this purpose, we will use the minkowskian
null coordinates $x^{\pm}=t\pm x$ in which
\[ds^2=-dt^2+dx^2=-dx^+dx^-\;.\]\\
After a Wick rotation $t=-i\tau$
\[ds^2=d\tau^2+dx^2=dzd\overline z\;,\]
where
\[\left\{\begin{array}{c}
z\\ \overline z\end{array}\right\}=\tau\pm ix=ix^{\pm}\;.\]\\\\
For weak curvatures, the matter action is
\[\begin{array}{lcl}
S\left[\eta_{ab}+h_{ab}\right]&=&S_0\left[\eta_{ab}\right]-
{1\over2\pi}\int d^2z\,T_{ab}\,h^{ab}\\\\
&=&{1\over2\pi}\int d^2z\,{1\over2}\,{\left(\partial f\right)}^2
-{1\over2\pi}\int d^2z\,T_{ab}\,h^{ab}\;.
\end{array}\]\\\\
The vacuum expectation value of the stress-energy tensor is
\[\begin{array}{lcl}
\left<T_{ab}\left(z,\overline z\right)\right>&=&\frac
{\int\left[df\right]\,e^{-S}\,T_{ab}\left(z,\overline z\right)}
{\int\left[df\right]e^{-S}}\\ 
&=&2\pi{1\over {Z\left[h^{ab}\right]}}\frac{\delta Z\left[h^{ab}\right]}
{\delta h^{ab}\left(z,\overline z\right)}=
2\pi\frac{\delta W\left[h^{ab}\right]}{\delta h^{ab}\left(z,\overline z\right)}
\;,
\end{array}\]
where we have defined
\[Z\left[h^{ab}\right]=\int\left[df\right]e^{-S}\equiv Z\left[0\right]
e^{W\left[h^{ab}\right]}\;.\]\\\\
$Z\left[h^{ab}\right]$ generates the correlation functions involving
stress-energy tensors of the free theory defined by $S_0$:
\[{\left<T_{a_1b_1}\left(z_1,\overline z_1\right)...
T_{a_nb_n}\left(z_n,\overline z_n\right)\right>}_0={\left(2\pi\right)}^n{\left.
\frac{\delta^n Z\left[h\right]}{\delta h^{a_1b_1}\left(z_1,\overline z_1\right)
...\delta h^{a_nb_n}\left(z_n,\overline z_n\right)}\right|}_{h=0}\;.\]\\\\
$W\left[h^{ab}\right]$ generates the corresponding connected correlators
\[{\left<T_{a_1b_1}\left(z_1,\overline z_1\right)...
T_{a_nb_n}\left(z_n,\overline z_n\right)\right>}_0^{\left(conn\right)}
={\left(2\pi\right)}^n{\left.
\frac{\delta^n W\left[h\right]}{\delta h^{a_1b_1}\left(z_1,\overline z_1\right)
...\delta h^{a_nb_n}\left(z_n,\overline z_n\right)}\right|}_{h=0}\;.\]
Since we will only use connected correlators, we will suppress the label 
$(conn)$ in what follows.\\\\
We study
\[\begin{array}{lcl}
\left<T_{ab}\left(z,\overline z\right)\right>-
{\left<T_{ab}\left(z,\overline z\right)\right>}_0&=&
2\pi\left({\left.\frac{\delta W}
{\delta h^{ab}\left(z,\overline z\right)}\right|}_h
-{\left.\frac{\delta W}
{\delta h^{ab}\left(z,\overline z\right)}\right|}_{h=0}\right)\\\\
\ &=&2\pi\int d^2z'
{\left.\frac{\delta^2 W}{\delta h^{ab}\left(z,\overline z\right)
\delta h^{cd}\left(z',\overline {z'}\right)}\right|}_{h=0}
h^{cd}\left(z',\overline {z'}\right)\\\\
\ &=&{1\over2\pi}\int d^2z'{\left<T_{ab}\left(z,\overline z\right)
T_{cd}\left(z',\overline {z'}\right)\right>}_0\,h^{cd}\left(z',\overline {z'}
\right)\;.\end{array}\]\\\\
In flat space
\[\left\{\begin{array}{ccccc}
T\left(z\right)&\equiv&T_{zz}\left(z\right)
&=&-{1\over2}:\partial_z f\partial_z f:\\\\
\overline T\left(\overline z\right)&\equiv&T_{\overline z\overline z}
\left(\overline z\right)
&=&-{1\over2}:\partial_{\overline z}f\partial_{\overline z}f:\\\\
T_{z\overline z}&=&0\;.\ &\ 
\end{array}\right.\]\\\\
As operators, $T$ and $\overline T$ depend only on $z$ and $\overline z$, 
respectively. This follows from the fact that $f\left(z,\overline z\right)$ is a
free field:
\[\partial_z\partial_{\overline z}f\left(z,\overline z\right)=0\;,\]
so
\[f\left(z,\overline z\right)=f_1\left(z\right)+f_2\left(\overline
z\right)\;.\]\\\\
The free $(h = 0)$ Euclidean matter action is
\[S_0\left[\eta_{ab}\right]={1\over2\pi}\int d^2z\,{1\over2}\,
{\left(\partial f\right)}^2\;,\]
so that the propagator satisfies
\[-\nabla^2G\left(z,z'\right)=2\pi\,\delta^{\left(2\right)}\left(z-z'\right)\;,
\]
with solution
\[G\left(z,\overline{z};z',\overline{z'}\right)=-\ln{\left|z-z'\right|}=
-{1\over2}\ln{\left(z-z'\right)
\left(\overline z-\overline {z'}\right)}\;.\]\\\\
Since ${\left<f\,f\right>}_0=G$, we
get
\[\left\{\begin{array}{ccc}
{\left<\partial_z f\partial_{z'}f\right>}_0&=&
-{1\over2}\,{1\over{\left(z-z'\right)}^2}\\\\
{\left<\partial_{\overline z}f\partial_{\overline {z'}}f\right>}_0&=&
-{1\over2}\,{1\over{\left(\overline z-\overline {z'}\right)}^2}\\\\
{\left<\partial_z f\partial_{\overline {z'}}f\right>}_0&=&0\;.
\end{array}\right.\]\\\\
By Wick theorem
\begin{equation}
\label{eq:2p}
\left\{\begin{array}{lcl}
{\left<T\left(z\right)T\left(z'\right)\right>}_0
&=&{1\over4}\,{\left<:\partial_z f\partial_z f:\;:\partial_{z'} f\partial_{z'} f:
\right>}_0={1\over4}\,2\,
{\left({\left<\partial_z f\partial_{z'}f\right>}_0\right)}^2\\\\
\ &=&{1\over8}\,{1\over{\left(z-z'\right)}^4}\\\\
{\left<\overline T\left(\overline z\right)
\overline T\left(\overline {z'}\right)\right>}_0&=&
{1\over8}\,{1\over{\left(\overline z-\overline {z'}\right)}^4}\\\\
{\left<T\left(z\right)\overline T\left(\overline {z'}\right)\right>}_0&=&0\;.
\end{array}\right.
\end{equation}\\\\
We get
\[\begin{array}{lcl}
\left<T\left(z,\overline z\right)\right>-{\left<T\left(z\right)\right>}_0=
\left<T\left(z,\overline z\right)\right>
&=&{1\over2\pi}\int d^2z'\,{\left<T\left(z\right)T\left(z'\right)\right>}_0\,
h^{zz}\left(z',\overline {z'}\right)\\\\
\ &=&{1\over16\pi}\int d^2z'\,\frac{h^{zz}\left(z',\overline {z'}\right)}
{{\left(z-z'\right)}^4}\,\Theta\left({\left|z-z'\right|}^2-a^2\right)\;.
\end{array}\]
Here, we have inserted a step function short-distance cutoff to render
the integral finite and we have used that $T$ is normal ordered in flat space
so that ${\left<T\right>}_0=0$.\\\\
This cutoff will introduce an explicit $\overline z$
-dependence in $\left<T\left(z,\overline z\right)\right>$:
\[\partial_{\overline z}\left<T\left(z,\overline z\right)\right>=
{1\over16\pi}\int d^2z'\,\frac{h^{zz}\left(z',\overline {z'}\right)}
{{\left(z-z'\right)}^3}\,\delta\left({\left|z-z'\right|}^2-a^2\right)\;.\]\\\\
We Taylor expand $h^{zz}\left(z',\overline {z'}\right)$ around
$\left(z,\overline z\right)$
\[h^{zz}\left(z',\overline {z'}\right)=h^{zz}\left(z,\overline {z}\right)+
\left(z'-z\right)\partial_z h^{zz}\left(z,\overline {z}\right)+
\left(\overline {z'}-\overline z\right)\partial_{\overline z}
h^{zz}\left(z,\overline z\right)+...\;.\]\\
Only the term ${\left(z'-z\right)}^3\,{1\over{3!}}\,\partial_z^3
h^{zz}\left(z,\overline {z}\right) $ in the expansion contributes to the
integral after the limit $a\rightarrow0$ has been taken
\[\begin{array}{lcl}
\partial_{\overline z}\left<T\left(z,\overline z\right)\right>
&=&-{1\over16\pi}\,{1\over{3!}}\,\partial_z^3h^{zz}\left(z,\overline {z}\right)
\int_0^{+\infty}{1\over2}\,dr^2\delta\left(r^2-a^2\right)\int_0^{2\pi}d\theta
\\\\
\ &=&-{1\over96}\,\partial_z^3h^{zz}\left(z,\overline {z}\right)\;.
\end{array}\]\\\\
Also
\[\partial_z\left<\overline T\left(z,\overline z\right)\right>=
-{1\over96}\,\partial_{\overline z}^{\,3}\,h^{\overline z\overline z}
\left(z,\overline {z}\right)\;.\]\\\\
The stress-energy tensor must obey the following three physical principles:\\\\
{\em First}, energy-momentum conservation requires a
divergenceless~stress-energy tensor:
\begin{eqnarray}
\label{eq:cons1}
\partial_{\overline z}\left<T\left(z,\overline z\right)\right>+
\partial_z\left<T_{\overline z z}\left(z,\overline z\right)\right>=0\\
\label{eq:cons2}
\partial_z\left<\overline T\left(z,\overline z\right)\right>+
\partial_{\overline z}\left<T_{z\overline z}\left(z,\overline z\right)\right>=0
\;.
\end{eqnarray}\\\\
{\em Second}, its trace
\[\begin{array}{lcl}
g^{ab}\left<T_{ab}\right>&=&\left(\eta^{ab}+h^{ab}\right)\left<T_{ab}\right>=
2\eta^{z\overline z}\left<T_{z\overline z}\right>+\cal O\left(h^2\right)\\\\\
&=&4\left<T_{z\overline z}\right>+\cal O\left(h^2\right)\;,
\end{array}\]
should be invariant under diffeomorphisms:
\[\delta_{\xi}h^{ab}\left(z,\overline {z}\right)=
\partial^a\xi^b\left(z,\overline {z}\right)+
\partial^b\xi^a\left(z,\overline {z}\right)\;.\]\\
{\em Finally}, it must be symmetric.\\\\
It is equivalent to require diffeomorphism
invariance of the effective action $W\left[h\right]$.\\\\
So we try to add local counterterms in $T_{ab}$ so as to respect these physical
principles.~(\ref{eq:cons1}) gives
\[\left<T_{\overline z z}\left(z,\overline z\right)\right>=
{1\over96}\partial_z^{\,2}h^{zz}\left(z,\overline {z}\right)\;,\]
and~(\ref{eq:cons2}) gives
\[\left<T_{z\overline z}\left(z,\overline z\right)\right>=
{1\over96}\partial_{\overline z}^{\,2}\,h^{\overline z\overline z}
\left(z,\overline {z}\right)\;.\]\\\\
Symmetry requires
\[\left<T_{\overline z z}\left(z,\overline z\right)\right>=
\left<T_{z\overline z}\left(z,\overline z\right)\right>=
{1\over96}\left(\partial_z^{\,2}h^{zz}\left(z,\overline {z}\right)+
\partial_{\overline z}^{\,2}\,h^{\overline z\overline z}
\left(z,\overline {z}\right)\right)\;,\]
which in turn implies
\[\begin{array}{lcl}
\left<T\left(z,\overline z\right)\right>&=&
-{1\over96}\partial_{\overline z}^{-1}
\partial_z^{\,3}h^{zz}\left(z,\overline {z}\right)
-{1\over96}\partial_{\overline z}\,\partial_zh^{\overline z\overline z}
\left(z,\overline {z}\right)\\\\
\ &=&-{1\over96}\partial_{\overline z}^{-1}\partial_z\left(
\partial_z^{\,2}h^{zz}\left(z,\overline {z}\right)+
\partial_{\overline z}^{\,2}\,h^{\overline z\overline z}
\left(z,\overline {z}\right)\right)\;,\end{array}\]
and
\[\left<\overline T\left(z,\overline z\right)\right>=
-{1\over96}\partial_z^{-1}\partial_{\overline z}\left(
\partial_{\overline z}^{\,2}h^{\overline z\overline z}
\left(z,\overline {z}\right)+\partial_z^{\,2}h^{zz}\left(z,\overline {z}\right)
\right)\;.\]\\\\
$T_{z\overline z}$ is not diffeomorphism invariant because
\[\delta_{\xi}\left<T_{z\overline z}\left(z,\overline z\right)\right>=
{1\over96}\left(2\,\partial_z^{\,2}\partial^z\xi^z\left(z,\overline z\right)+
2\,\partial_{\overline z}^{\,2}\partial^{\overline z}\xi^{\overline z}
\left(z,\overline z\right)\right)\;.\]\\\\
However
\[\begin{array}{lcl}
\delta_{\xi}\left(\partial_{\overline z}\partial_z
h^{\overline z z}\left(z,\overline {z}\right)\right)&=&
\partial_{\overline z}\partial_z\partial^{\overline z}\xi^z
\left(z,\overline {z}\right)+\partial_{\overline z}\partial_z\partial^{z}
\xi^{\overline z}\left(z,\overline z\right)\\\\
\ &=&\partial_z^{\,2}\partial^z\xi^z\left(z,\overline z\right)+
\partial_{\overline z}^{\,2}\partial^{\overline z}\xi^{\overline z}
\left(z,\overline z\right)\;,
\end{array}\]
so add
\[{1\over96}\left(-2\right)\partial_{\overline z}\partial_z
h^{\overline z z}\left(z,\overline {z}\right)\;,\]
to $\left<T_{z\overline z}\right>$
\[\left<T_{z\overline z}\left(z,\overline z\right)\right>=
{1\over96}\left(\partial_z^{\,2}h^{zz}\left(z,\overline {z}\right)-
2\,\partial_z\partial_{\overline z}\,h^{z\overline z}
\left(z,\overline {z}\right)+
\partial_{\overline z}^{\,2}\,h^{\overline z\overline z}
\left(z,\overline {z}\right)\right)\;,\]
which implies
\[\left<T\left(z,\overline z\right)\right>=-{1\over96}
\partial_{\overline z}^{-1}\partial_z
\left(\partial_z^{\,2}h^{zz}\left(z,\overline {z}\right)-
2\,\partial_z\partial_{\overline z}\,h^{z\overline z}
\left(z,\overline {z}\right)+
\partial_{\overline z}^{\,2}\,h^{\overline z\overline z}
\left(z,\overline {z}\right)\right)\;,\]
and
\[\left<\overline T\left(z,\overline z\right)\right>=-{1\over96}
\partial_z^{-1}\partial_{\overline z}
\left(\partial_z^{\,2}h^{zz}\left(z,\overline {z}\right)-
2\,\partial_z\partial_{\overline z}\,h^{z\overline z}
\left(z,\overline {z}\right)+
\partial_{\overline z}^{\,2}\,h^{\overline z\overline z}
\left(z,\overline {z}\right)\right)\;.\]\\\\
The trace is
\[g^{ab}\left<T_{ab}\right>=4\left<T_{z\overline z}\right>=
{1\over24}\left(\partial_z^{\,2}h^{zz}-
2\,\partial_z\partial_{\overline z}\,h^{z\overline z}+
\partial_{\overline z}^{\,2}\,h^{\overline z\overline z}\right)\;.\]\\\\
One recognizes the right hand side of the above equation as the curvature scalar
to first order in $h$
\[\begin{array}{lcl}
\delta R&=&g^{ab}\delta R_{ab}=g_{cd}\nabla^2\delta g^{cd}-
\nabla_a\nabla_b\delta g^{ab}\\\\
\ &=&\eta_{cd}\,\eta^{ef}\partial_e\partial_f h^{cd}
-\partial_a\partial_bh^{ab}\\\\
\ &=&4\,\eta_{z\overline z}\,\eta^{z\overline z}\partial_z\partial_{\overline z}
h^{z\overline z}-\partial_z^2h^{zz}-2\,\partial_z\partial_{\overline z}
h^{z\overline z}-\partial_{\overline z}^2h^{\overline z\overline z}\\\\
\ &=&-\left(\partial_z^{\,2}h^{zz}-
2\,\partial_z\partial_{\overline z}\,h^{z\overline z}+
\partial_{\overline z}^{\,2}\,h^{\overline z\overline
z}\right)\;.\end{array}\]\\\\
So
\[\left\{\begin{array}{ccc}
\left<T_{z\overline z}\right>&=&-{1\over24}\,R\\\\
\left<T\right>&=&{1\over96}\,\partial_{\overline z}^{-1}\partial_z\,R\\\\
\left<\overline T\right>&=&{1\over96}\,\partial_z^{-1}\partial_{\overline z}\,R
\;.
\end{array}\right.\]\\\\
Integrating
\[\left<T_{ab}\right>=2\pi\frac{\delta W\left[h\right]}{\delta h^{ab}}\;,\] 
we find
\[\begin{array}{lcl}
W\left[h\right]&=&{1\over2\pi}\left(-{1\over48}\right)\int d^2z
\left(\partial_z^{\,2}h^{zz}-
2\,\partial_z\partial_{\overline z}\,h^{z\overline z}+
\partial_{\overline z}^{\,2}\,h^{\overline z\overline z}\right)
{\left(4\partial_z\partial_{\overline z}\right)}^{-1}\\\\
\ &.&\left(\partial_z^{\,2}h^{zz}-
2\,\partial_z\partial_{\overline z}\,h^{z\overline z}+
\partial_{\overline z}^{\,2}\,h^{\overline z\overline z}\right)\\\\
\ &=&{1\over2\pi}\left(-{1\over48}\right)\int d^2z\,
R\,{\left(\nabla^2\right)}^{-1}R\;.\end{array}\]
This effective action is called the Polyakov action (it does
not contain any term linear in $h$ because ${\left<T\right>}_0=0$).\\\\
These relations also hold in arbitrary spacetimes \cite{Pol}, i.e.\,,
the curvature need not
be weak, as we assumed in the derivation above. Thus
\begin{equation}
g^{ab}\left<T_{ab}\right>=+{1\over24}\,R\;,
\end{equation}
and
\begin{equation}
W\left[g^{ab}\right]={1\over2\pi}\left(+{1\over48}\right)\int d^2x
\,\sqrt{-g}\,R{\left(\nabla^2\right)}^{-1}R\;.
\end{equation}\\\\
In a general spacetime, we can evaluate the stress-energy tensor induced
by the Polyakov action in the conformal gauge: one way is to vary the
Polyakov action and then to go to conformal gauge. The calculation being quite
lengthy, we will proceed as follows: we first evaluate $T_{+-}$ using the
conformal anomaly:
\[T_{+-}={\left(2\,g^{+-}\right)}^{-1}\left(+{1\over24}\right)R=-{1\over12}\,
\partial_+\partial_-\rho\;.\]\\
Then, we determine $T_{\pm\pm}$ by requiring $T$ to be divergenceless:
\[\begin{array}{llll}
\ &\nabla_{\mp}T_{\pm\pm}&+&\nabla_{\pm}T_{\mp\pm}=0\\\\
\Leftrightarrow&\partial_{\mp}T_{\pm\pm}&+&\left(\partial_{\pm}-
\Gamma^{\pm}_{\pm\pm}\right)T_{\mp\pm}=0\\\\
\Leftrightarrow&\partial_{\mp}T_{\pm\pm}&=&
{1\over12}\left(\partial_{\mp}\partial_{\pm}^2\rho-
2\partial_{\pm}\rho\partial_{\pm}\partial_{\mp}\rho\right)\\\\
\ &\ &=&{1\over12}\partial_{\mp}\left(\partial_{\pm}^2\rho-
\partial_{\pm}\rho\partial_{\pm}\rho\right)\;.\end{array}\]\\\\
So
\[T_{\pm\pm}={1\over12}\left(\partial_{\pm}^2\rho-
\partial_{\pm}\rho\partial_{\pm}\rho+t_{\pm}\left(x^{\pm}\right)\right)
\;.\]\\\\
The functions $t_{\pm}\left(x^{\pm}\right)$ are determined by boundary
conditions. They reflect the non-locality of the Polyakov action.\\\\
Following \cite{CGHS}, we compute this quantum induced stress-energy tensor for
the collapsing two
dimensional dilaton black hole displayed in fig.~\ref{fig:2Dcoll2}, with
metric
given by (\ref{eq:2Dcoll}) in Kruskal coordinates. In the asymptotically
minkowskian coordinates $\sigma^{\pm}$, covering the region outside the black
hole in fig.~\ref{fig:2Dcoll2}, the metric takes the form
\begin{equation}
\label{eq:met}
e^{-2\rho'}=\left\{\begin{array}{lcl}
1+{M\over{\lambda^2x^+_0}}e^{\lambda\sigma^-}&,&x^+<x^+_0\nonumber\\\\
1+{M\over\lambda}e^{-2\lambda\sigma}&,&x^+>x^+_0\;.
\end{array}\right.
\end{equation}\\\\
The coordinates $\sigma^{\pm}$ are thus not minkowskian in the (flat) linear
dilaton vacuum, $x^+<x^+_0$.\\\\
In terms of $\sigma^{\pm}$,
the stress-energy tensor is
\begin{equation}
\label{eq:stress}
\left\{\begin{array}{lll}
T_{\sigma^{\pm}\sigma^{\pm}}&=&{1\over12}\left(\partial_{\sigma^{\pm}}^2\rho'-
\partial_{\sigma^{\pm}}\rho'\partial_{\sigma^{\pm}}\rho'+
t_{\sigma^{\pm}}\left(\sigma^{\pm}\right)\right)\\\\
\ &=&-{1\over24}\left(e^{2\rho'}\partial_{\sigma^{\pm}}^2e^{-2\rho'}-
{1\over2}e^{4\rho'}{\left(\partial_{\sigma^{\pm}}e^{-2\rho'}\right)}^2-
2t_{\sigma^{\pm}}\left(\sigma^{\pm}\right)\right)\\\\
T_{\sigma^+\sigma^-}&=&-{1\over12}\partial_{\sigma^+}\partial_{\sigma^-}\rho'=
{1\over24}\left(e^{2\rho'}\partial_{\sigma^+}\partial_{\sigma^-}e^{-2\rho'}-
e^{4\rho'}{\left(\partial_{\sigma^+}e^{-2\rho'}\right)}
{\left(\partial_{\sigma^-}e^{-2\rho'}\right)}\right)\;.
\end{array}\right.
\end{equation}\\\\
Since the coordinates $\sigma^{\pm}$ are minkowskian at $J^{\pm}_R$\,, $\rho'=0$
there, and
\[\left\{\begin{array}{cc}
T_{\sigma^{\pm}\sigma^{\pm}}=&{1\over12}\,t_{\sigma^{\pm}}
\left(\sigma^{\pm}\right)\\\\
T_{\sigma^+\sigma^-}=&0\;.
\end{array}\right.\]\\\\
If the incoming quantum state of the matter $f$ is the vacuum state
$\left|0_{in}\right>$, then $T$,
which should be written $\left<0_{in}|T|0_{in}\right>$, vanishes at $J^-_R$,
hence
\[\begin{array}{lcl}
t_{\sigma^+}\left(\sigma^+\right)&=&0\\\\
t_{\sigma^-}\left(\sigma^-=-\infty\right)&=&0\;.
\end{array}\]\\\\
$T$ must also vanish in the linear dilaton region, $x^+<x^+_0$.
Inserting~(\ref{eq:met}) in~(\ref{eq:stress}) we obtain
\[\begin{array}{lll}
0=T_{\sigma^+\sigma^+}={1\over12}\,t_{\sigma^+}\left(\sigma^+\right)=0
&,&x^+<x^+_0\\\\
0=T_{\sigma^+\sigma^-}=0&,&x^+<x^+_0\;,
\end{array}\]
and
\[\begin{array}{lll}
0=T_{\sigma^-\sigma^-}&=&-{1\over24}\left\{{\left(1+{M\over{\lambda^2x_0^+}}
e^{\lambda\sigma^-}\right)}^{-1}{M\over x_0^+}e^{\lambda\sigma^-}-{1\over2}
{\left(1+{M\over{\lambda^2x_0^+}}e^{\lambda\sigma^-}
\right)}^{-2}\right.\\
\ &\,*&\left.{\left({M\over{\lambda x_0^+}}e^{\lambda\sigma^-}\right)}^2
-2t_{\sigma^-}\left(\sigma^-\right)\right\}\\\\
\ &=&-{1\over24}\left\{-{1\over2}\left[{\left(\frac
{\left({M\over{\lambda x_0^+}}e^{\lambda\sigma^-}\right)}
{\left(1+{M\over{\lambda^2x_0^+}}e^{\lambda\sigma^-}\right)}-
\lambda\right)}^2-\lambda^2\right]-
2t_{\sigma^-}\left(\sigma^-\right)\right\}\\\\
\ &=&-{1\over24}\left\{{\lambda^2\over2}\left[1-{\left(1+
{M\over{\lambda^2x_0^+}}e^{\lambda\sigma^-}\right)}^{-2}\right]-
2t_{\sigma^-}\left(\sigma^-\right)\right\}\\\\
\ &\Leftrightarrow&t_{\sigma^-}\left(\sigma^-\right)={\lambda^2\over4}
\left(1-{\left(1+{M\over{\lambda^2\,x^+_0}}
e^{\lambda\sigma^-}\right)}^{-2}\right)\;.
\end{array}\]\\\\
The stress-energy tensor is now completely determined; at $J^+_{R}$, it is
equal to
\[\left\{\begin{array}{l}
T_{\sigma^+\sigma^-}=0\\\\
T_{\sigma^+\sigma^+}=0\\\\
T_{\sigma^-\sigma^-}={1\over12}\,t_{\sigma^-}\left(\sigma^-\right)=
{\lambda^2\over48}\left(1-{\left(1+{M\over{\lambda^2x^+_0}}e^{\lambda\sigma^-}
\right)}^{-2}\right)\;.
\end{array}\right.\]\\\\
This represents an energy flow in the $\sigma^+$-direction equal to 
$T_{\sigma^-\sigma^-}$ per unit length in the $\sigma^-$-direction. In the far
past, this flux is zero but it builds up to the constant value
${\lambda^2\over48}$ as the horizon is approached: this is the Hawking
radiation. We note that the flux is independent of the mass of the black hole
unlike the four dimensional Schwarzschild black hole where the flux is found to
be inversely proportional to the square of the mass.\\\\
As we have shown in section 3, the temperature of a black hole
is ${\kappa\over2\pi}$, where $\kappa$ is the surface gravity of the black hole
defined by
\[\kappa=\lim_H\sqrt{{\left(\nabla V\right)}^2}\;,\]
where $V^2$ is the redshift factor
\[V^2=-\xi_a\xi^a\;,\]
$\xi^a$ being the time translation Killing field of the black hole.\\\\
The
metric of the eternal two dimensional black hole of fig.~\ref{fig:2DSch} was
given by:
\[ds^2={\left(1+{M\over\lambda}e^{-2\lambda\sigma}\right)}^{-1}
\left(-d\tau^2+d\sigma^2\right)\;.\]\\
Thus, $\xi^a={\left(\frac{\partial}{\partial\tau}\right)}^a$, as we showed
before, and
\[V^2=-g_{ab}\,\xi^a\xi^b=-g_{\tau\tau}=
{\left(1+{M\over\lambda}e^{-2\lambda\sigma}\right)}^{-1}\;.\]\\
Then
\begin{eqnarray}
\kappa&=&\lim_{\sigma\rightarrow-\infty}\sqrt{g^{\sigma\sigma}
{\left(\partial_{\sigma}V\right)}^2}=
\lim_{\sigma\rightarrow-\infty}\sqrt{g^{\sigma\sigma}}
\left|\partial_{\sigma}V\right|\nonumber\\
\ &=&\lim_{\sigma\rightarrow-\infty}{1\over\sqrt{g_{\sigma\sigma}}}{1\over2V}
\left|\partial_{\sigma}V^2\right|=\lim_{\sigma\rightarrow-\infty}{1\over2V^2}
\left|\partial_{\sigma}V^2\right|\nonumber\\
\ &=&\lim_{\sigma\rightarrow-\infty}
{1\over2}\left(1+{M\over\lambda}e^{-2\lambda\sigma}\right)
\left|-{\left(1+{M\over\lambda}e^{-2\lambda\sigma}\right)}^{-2}{M\over\lambda}
\left(-2\lambda\right)e^{-2\lambda\sigma}\right|\nonumber\\
\ &=&\lim_{\sigma\rightarrow-\infty}
{\left(1+{M\over\lambda}e^{-2\lambda\sigma}\right)}^{-1}
{M\over\lambda}e^{-2\lambda\sigma}\lambda=\lambda\;.
\end{eqnarray}\\\\
So the temperature of the two dimensional dilaton black hole is
${\lambda\over{2\pi}}$.\\\\
We can check that the outgoing radiation is thermal by canonically quantizing
the matter field $f$ in the classical background geometry of the collapsing two
dimensional dilaton black hole depicted in fig.~\ref{fig:2Dcoll2}. Here, we
closely follow the corresponding derivation for a Schwarzschild black hole (For
more details, see \cite{GN}).
\\\\The coordinates $\sigma^{\pm}$ defined by
\[\left\{\begin{array}{cclcl}
\lambda x^+&=&e^{\lambda\sigma^+}&\ &\ \\\\
\lambda\left(x^-+\Delta\right)&=&-e^{-\lambda\sigma^-}&,&
\Delta={M\over{\lambda^3x^+_0}}\;,
\end{array}\right.\]
were minkowskian at $J^{\pm}_R$.\\\\
On the other hand, the coordinates
\[
\left\{\begin{array}{c}
\lambda x^+=e^{\lambda y^+}\\\\
\lambda x^-=-\Delta\lambda e^{-\lambda y^-}\;,
\end{array}\right.
\]
are minkowskian in the linear dilaton region and in
particular at $J^-_L$.\\\\
The transformations between $y^-$ and $\sigma^-$ are 
\begin{equation}
\label{eq:trans}
\left\{\begin{array}{l}
\sigma^-=-{1\over\lambda}\ln{\left(\lambda\Delta\left(e^{-\lambda y^-}
-1\right)\right)}\\\\
y^-=-{1\over\lambda}\ln{\left({1\over{\lambda\Delta}}e^{-\lambda\sigma^-}
+1\right)}\;,
\end{array}\right.
\end{equation}
and the horizon is located at $y^-=0$.\\\\
In the present case, there is no back-scattering since $f$ is a free field:
\[\partial_+\partial_- f=0\;.\]\\\\
We concentrate on the right moving field modes because they are the ones
responsible for the Hawking effect at $J^+_R$.\\\\
The modes defined by
\[v_{\omega}\left(\sigma^-\right)={1\over\sqrt{2\omega}}\,e^{-i\omega\sigma^-}
\;,\]
appear to be positive frequency at $J^+_R$.
The corresponding wave packets
\[v_{jn}\left(\sigma^-\right)={1\over\sqrt{E}}\int_{jE}^{\left(j+1\right)E}
d\omega\,e^{\frac{2\pi i\omega n}{E}}v_{\omega}\left(\sigma^-\right)\;,\]
have frequencies around $\left(j+{1\over2}\right)E$ and are peaked around
$u=\frac{2\pi n}{E}$ with time spread $\Delta u={{2\pi}\over E}$.\\\\
Propagating
$v_{\omega}\left(\sigma^-\right)$ back to $J^-_L$ and expressing it as a
function of $y^-$, we get
\[v_{\omega}\left(\sigma^-\left(y^-\right)\right)={1\over\sqrt{2\omega}}\,
e^{{{i\omega}\over\lambda}\ln{\left(\lambda\Delta\left(e^{-\lambda y^-}-
1\right)\right)}}\Theta\left(-y^-\right)\;,\]
and close to the horizon, i.e.\,, at late times $\sigma^-$,
\[v_{\omega}\left(\sigma^-\left(y^-\right)\right)\approx
{1\over\sqrt{2\omega}}\,
e^{{{i\omega}\over\lambda}\ln{\left(-\lambda^2\Delta y^-\right)}}\ ,\hspace{1cm}
0<-\lambda y^-\ll 1\;.\]\\\\
Relative to $J^-_L$, the positive frequency modes are
\[u_{\omega}\left(y^-\right)={1\over\sqrt{2\omega}}\,e^{-i\omega y^-}\;,\]
and $v_{\omega}\left(y^-\right)$ and $v_{jn}\left(y^-\right)$ do not appear to
be purely positive frequency relative to $J^-_L$: it can be shown that for late
times, i.e.\,, for large n,
\[\hat{v}_{jn}\left(-\omega'\right)=-e^{-{{\pi\omega_j}\over\lambda}}
\hat{v}_{jn}\left(\omega'\right)\ ,\hspace{1cm}\omega'>0\;,\]
where $\hat{v}_{jn}\left(\omega'\right)$ is defined by
\[v_{jn}\left(y^-\right)=\int_0^{+\infty}d\omega'\left(
\hat{v}_{jn}\left(\omega'\right)u_{\omega'}\left(y^-\right)+
\hat{v}_{jn}\left(-\omega'\right)u_{\omega'}^*\left(y^-\right)\right)\;.\]\\\\
Proceeding in the same way as for the Schwarzschild black hole, we can get
the density matrix describing measurements at $J^+_R$ on a given late time mode
$\left(i\right)\equiv\left(jn\right)$:       
\[D_i=\sum_{N=0}^{+\infty}e^{-N{{2\pi\omega_i}\over\lambda}}
\left|v_i^N\right>\left<v_i^N\right|\;.\]\\
This is a completely thermal density matrix with temperature
${\lambda\over{2\pi}}$.\\\\
We now calculate
\[\begin{array}{llll}
\ &\left<\psi\right|T^{\left(out\right)}_{\sigma^-\sigma^-}
\left(\sigma^-\right)\left|\psi\right>&=&\left<0_{in}\right|S^{-1}\,
T^{\left(out\right)}_{\sigma^-\sigma^-}\left(\sigma^-\right)
S\left|0_{in}\right>\\\\
=&\left<0_{in}\right|T^{\left(in\right)}_{\sigma^-\sigma^-}\left(\sigma^-\right)
\left|0_{in}\right>&=&\left<0_{in}\right|{1\over2}\,\partial_{\sigma^-}
f^{\left(in\right)}\left(\sigma^-\right)
\partial_{\sigma^-}f^{\left(in\right)}\left(\sigma^-\right)\left|0_{in}\right>
\;,\end{array}\]
at $J^+_R$ .\\\\
${\left<T^{\left(in\right)}_{\sigma^-\sigma^-}
\left(\sigma^-\right)\right>}_{in}$ being divergent, we regularize it by the 
point splitting method, thus we define:
\[{\left<T^{\left(in\right)}_{\sigma^-\sigma^-}
\left(\sigma^-\right)\right>}_{in}\equiv\lim_{\delta\rightarrow0}
{\left<{1\over2}\,\partial_{\sigma^-}
f^{\left(in\right)}\left(\sigma^--{1\over2}\,\delta\right)
\partial_{\sigma^-}f^{\left(in\right)}\left(\sigma^-+{1\over2}\,\delta\right)
\right>}_{in}\;.\]\\\\
Now
\[f^{\left(in\right)}\left(\sigma^-,\sigma^+\right)=\int_{0}^{+\infty}d\omega
\left\{a^{\left(in\right)}_{\omega}
u_{\omega}\left(y^-\left(\sigma^-\right)\right)+
{\left(a^{\left(in\right)}_{\omega}\right)}^{\dagger}
u_{\omega}^*\left(y^-\left(\sigma^-\right)\right)+...\right\}\;,\]
where the ellipsis stand for the $\sigma^+$-dependence of $f$.\\\\
Thus
\[\begin{array}{lcl}
{\left<T^{\left(in\right)}_{\sigma^-\sigma^-}\left(\sigma^-\right)\right>}_{in}
&=&{1\over4}\,\int_{0}^{+\infty}{{d\omega}\over\omega}\,
\partial_{\sigma^-}u_{\omega}
\left(y^-\left(\sigma^--{1\over2}\,\delta\right)\right)\,
\partial_{\sigma^-}u_{\omega}^*
\left(y^-\left(\sigma^-+{1\over2}\,\delta\right)\right)\\\\
&=&{1\over4}\,{\left(\frac{dy^-}{d\sigma^-}\right)}_{\sigma^--{1\over2}\,\delta}
\;{\left(\frac{dy^-}{d\sigma^-}\right)}_{\sigma^-+{1\over2}\,\delta}\\
&.&\int_{0}^{+\infty}{{d\omega}\over\omega}\,
\frac{du_{\omega}}{dy^-}
\left(y^-\left(\sigma^--{1\over2}\,\delta\right)\right)
\,\frac{du_{\omega}^*}{dy^-}
\left(y^-\left(\sigma^-+{1\over2}\,\delta\right)\right)\\\\
&=&{1\over4}\,{\left(\frac{dy^-}{d\sigma^-}\right)}_{\sigma^--{1\over2}\,\delta}
\;{\left(\frac{dy^-}{d\sigma^-}\right)}_{\sigma^-+{1\over2}\,\delta}\\
&.&\int_{0}^{+\infty}d\omega\,\omega\,
u_{\omega}\left(y^-\left(\sigma^--{1\over2}\,\delta\right)\right)\,
u_{\omega}^*\left(y^-\left(\sigma^-+{1\over2}\,\delta\right)\right)\\\\
&=&{1\over4}\,{\left(\frac{dy^-}{d\sigma^-}\right)}_{\sigma^--{1\over2}\,\delta}
\;{\left(\frac{dy^-}{d\sigma^-}\right)}_{\sigma^-+{1\over2}\,\delta}\\
&.&\int_{0}^{+\infty}d\omega\,\omega\,
\exp{\left\{-i\omega\left[y^-\left(\sigma^--{1\over2}\,\delta\right)-
y^-\left(\sigma^-+{1\over2}\,\delta\right)\right]\right\}}\\\\
&=&{1\over4}\,{\left(\frac{dy^-}{d\sigma^-}\right)}_{\sigma^--{1\over2}\,\delta}
\;{\left(\frac{dy^-}{d\sigma^-}\right)}_{\sigma^-+{1\over2}\,\delta}\;\;
{\left[y^-\left(\sigma^--{1\over2}\,\delta\right)-
y^-\left(\sigma^-+{1\over2}\,\delta\right)\right]}^{-2}\;.
\end{array}\]\\\\
Taylor expanding around $\delta=0$ and taking the limit $\delta\rightarrow0$,
we obtain
\[{\left<T^{\left(in\right)}_{\sigma^-\sigma^-}
\left(\sigma^-\right)\right>}_{in}
\rightarrow-{1\over4\delta^2}-{1\over24}\left(\frac{y^{-^{'''}}}{y^{-^{'}}}
-{3\over2}{\left(\frac{y^{-^{''}}}{y^{-^{'}}}\right)}^2\right)\;,\]
where the primes denote derivatives with respect to $\sigma^-$.\\\\
The term inside the parentheses is the so called schwarzian derivative which
can be evaluated by using the explicit form of the transformation
$y^-\left(\sigma^-\right)$ given by~(\ref{eq:trans}). The result is
\[{\left<T^{\left(in\right)}_{\sigma^-\sigma^-}
\left(\sigma^-\right)\right>}_{in}
\rightarrow-{1\over{4\delta^2}}+{\lambda^2\over48}\left(1-
{1\over{{\left(1+\lambda\Delta e^{\lambda\sigma^-}\right)}^2}}\right)\;.\]\\
The divergent term we recognize as the expectation value of
$T^{\left(out\right)}_{\sigma^-\sigma^-}$ in the out-vacuum. Substracting this
term, we obtain
\begin{equation}
\label{eq:T}
{\left<T^{\left(out\right)}_{\sigma^-\sigma^-}
\left(\sigma^-\right)\right>}_{\psi}=
{\left<T^{\left(in\right)}_{\sigma^-\sigma^-}
\left(\sigma^-\right)\right>}_{in}=
{\lambda^2\over48}\left(1-
{1\over{{\left(1+\lambda\Delta e^{\lambda\sigma^-}\right)}^2}}\right)\;.
\end{equation}\\\\
This agrees with the earlier calculation based on the trace anomaly.
Furthermore, it
shows that the quantum induced stress-energy tensor really arises from
the Hawking radiation.\\\\
The above result can also be derived by using the conformal properties of the
quantized scalar field $f$ (conformal field theory is reviewed in
\cite{Ginsparg} and \cite{Cardy}).\\
First, the two-point correlators (\ref{eq:2p})
of the stress-energy tensor
defines the central charge associated with $f$; here $c=1$.
\\Second, it is known
that the stress-energy tensor transforms with an anomalous term proportional
to $c$ times the schwarzian derivative under a conformal transformation (that
is, a transformation which preserves the conformal gauge)
\begin{equation}
\label{eq:transT}
T_{\sigma^-\sigma^-}\left(\sigma^-\right)=
{\left(\frac{dy^-}{d\sigma^-}\right)}^2T_{y^-y^-}\left(y^-\right)-
{c\over24}D_{\sigma^-}^S\left(y^-\right)\;,
\end{equation}
where $D_{\sigma^-}^S\left(y^-\right)$ is the schwarzian derivative defined
above
\[D_{\sigma^-}^S\left(y^-\right)=\frac{y^{-^{'''}}}{y^{-^{'}}}
-{3\over2}{\left(\frac{y^{-^{''}}}{y^{-^{'}}}\right)}^2\;.\]\\\\
In the derivation of this anomalous transformation law \cite[pages
182-183]{Cardy},
$T_{y^-y^-}\left(y^-\right)$ is assumed to be normal-ordered , i.e.\,,  
${\left<T_{y^-y^-}^{\left(in\right)}\left(y^-\right)\right>}_{in}=0$; taking the
expectation value of (\ref{eq:transT}), we get
\[{\left<T_{\sigma^-\sigma^-}^{\left(in\right)}
\left(\sigma^-\right)\right>}_{in}=
0-{1\over24}D_{\sigma^-}^S\left(y^-\right)={\lambda^2\over48}\left(1-
{1\over{{\left(1+\lambda\Delta e^{\lambda\sigma^-}\right)}^2}}\right)\;,\]
where we have set $c=1$. This agrees with (\ref{eq:T}).\\\\
Thus, ${\left<T_{\sigma^-\sigma^-}^{\left(in\right)}
\left(\sigma^-\right)\right>}_{in}$
cannot be normal-ordered; we now know that this is because of the Hawking
radiation:
\[{\left<T_{\sigma^-\sigma^-}^{\left(in\right)}
\left(\sigma^-\right)\right>}_{in}=
{\left<T_{\sigma^-\sigma^-}^{\left(out\right)}
\left(\sigma^-\right)\right>}_{\psi}\neq0\;.\]\\\\
Therefore, the three methods of calculating the stress-energy tensor arising
from the Hawking radiation give the same answer.

\newpage
\subsection{Back Reaction}
We now study the effect of the back reaction of the Hawking radiation
on the background geometry \cite{ST1,ST2}.\\\\
In the two dimensional "Einstein" equations,
we add the quantum induced stress-energy tensor to the right hand side.
In the conformal gauge, the Einstein equation for $\rho$ (or $g_{+-}$)
becomes
\[2\,e^{-2\Phi}\left(\partial_+\partial_-\Phi-
\left(2\,\partial_+\partial_-\Phi-2\,\partial_+\Phi\partial_-\Phi-
{1\over2}\lambda^2e^{2\rho}\right)\right)=-{N\over12}
\partial_+\partial_-\rho\;,\]\\
and the constraints, i.e.\,, the Einstein equations for $g_{\pm\pm}$, are
\[2\,e^{-2\Phi}\left(\partial_{\pm}^2\Phi-
2\,\partial_{\pm}\rho\partial_{\pm}\Phi\right)={1\over2}\sum_i
\partial_{\pm}f_i\partial_{\pm}f_i+{N\over12}\left(\partial_{\pm}^2\rho-
{\left(\partial_{\pm}\rho\right)}^2+t_{\pm}\left(x^{\pm}\right)\right)\;.\]\\\\
The equations for $\Phi$ and $f_i$ are unchanged
\[\partial_+\partial_-f_i=0\]
\[-\partial_+\partial_-\rho+2\,\partial_+\partial_-\Phi-
2\,\partial_+\Phi\partial_-\Phi-{1\over2}\lambda^2e^{2\rho}=0\;.\]\\\\
The following combinations of the equations for $\rho$ and $\Phi$ will be useful
\begin{eqnarray}
\label{eq:BR1}
\partial_+\partial_-\Phi&=&\left(1-{1\over24}Ne^{2\Phi}\right)
\partial_+\partial_-\rho\\
\label{eq:BR2}
2\left(1-{1\over12}Ne^{2\Phi}\right)\partial_+\partial_-\Phi&=&
\left(1-{1\over24}Ne^{2\Phi}\right)\left(4\,\partial_+\Phi\partial_-\Phi+
\lambda^2e^{2\rho}\right)\;.
\end{eqnarray}\\\\
As we see from the two dimensional Einstein equations, $e^{2\Phi}$ is the
(square of the) gravitational coupling strength and depends on position.
Classically, it varies from zero, asymptotically far away from the black hole,
to ${\lambda\over M}$, on the horizon. The semiclassical equations are valid
as long as $e^{2\Phi}$ is small.\\\\
The linear dilaton vacuum remains an exact solution of these semiclassical
equations. This is because the quantum induced stress-energy tensor is zero
since the curvature itself is zero.\\\\
We can try to find the spacetime due to an incoming
shell of matter moving with the velocity of light. In the interior of the shell,
the spacetime is flat and in the exterior, it is given by some solution of the
semiclassical equations. Unfortunately, these are nonlinear partial differential
equations which cannot be made linear, as was the case for the classical
equations (see (\ref{ClCGHS})), and for these reasons, they have never been
solved in closed form.\\\\
However, close
to the infall line, $x^+=x^+_0+\varepsilon$, the behaviour of the solution can
be found if we assume that it matches continuously onto the vacuum across
$x^+=x^+_0$ and that it approaches the classical black hole solution when
$x^-\rightarrow-\infty$.\\\\
In fact, $\Sigma\equiv\partial_+\Phi\left(x^+_0+\varepsilon,x^-\right)$ (which
is different from $\partial_+\Phi\left(x^+_0-\varepsilon,x^-\right)$, i.e.\,,
$\partial_+\Phi\left(x^+,x^-\right)$ is discontinuous at $x^+_0$)
satisfies (\ref{eq:BR2}) which becomes an ordinary differential equation
\[2\left(1-{1\over12}N\,e^{2\Phi}\right)\partial_-\Sigma-
4\left(1-{1\over24}N\,e^{2\Phi}\right)\partial_-\Phi\,\Sigma=
\left(1-{1\over24}N\,e^{2\Phi}\right)\lambda^2e^{2\rho}\;,\]
here
\[e^{-2\Phi}=e^{-2\rho}=-\lambda^2x^+_0x^-\;.\]\\\\
One solution of this inhomogeneous equation is a constant equal to
\[\partial_+\Phi\left(x^+_0-\varepsilon,x^-\right)=-{1\over{2 x^+_0}}\;.\]\\
The general solution can then be found by integration
\begin{equation}
\label{eq:Sigma}
\Sigma+{1\over{2 x^+_0}}=\frac{Ke^{2\Phi}}{\sqrt{\left(1-{N\over12}e^{2\Phi}
\right)}}=\frac{K}{\sqrt{-\lambda^2x^+_0x^-}\sqrt{-\lambda^2x^+_0x^--{N\over12}}
}\;.\end{equation}\\\\
As $x^-\rightarrow-\infty$,
\[\Sigma\sim-{1\over{2 x^+_0}}+\frac{K}{-\lambda^2x^+_0x^-}\;.\]\\
The classical solution has
\[\partial_+\Phi=-{1\over2}\frac{\partial_+e^{-2\Phi}}{e^{-2\Phi}}=
-{1\over{2 x^+_0}}-\frac{\Delta}{2x^+_0x^-}\;,\]
thus
\[K={1\over2}\lambda^2\Delta=\frac{M}{2\lambda x^+_0}\;.\]\\\\
$\Sigma$ is singular for
\[e^{-2\Phi}=-\lambda^2x^+_0x^-={N\over12}\;.\]\\
The singularity is a curvature singularity because
\[\begin{array}{lll}
R&=&8e^{-2\rho}\partial_+\partial_-\rho=8e^{-2\rho}
{\left(1-{1\over24}\,Ne^{2\Phi}\right)}^{-1}\partial_-\Sigma\\\\
\ &=&4{\left(1-{1\over12}\,Ne^{2\Phi}\right)}^{-1}
\left(4e^{-2\rho}\,\partial_-\Phi\,\Sigma+\lambda^2\right)\;,
\end{array}\]
where we have used~(\ref{eq:BR1}) and~(\ref{eq:BR2}). The singularity is behind
the classical horizon, where $e^{-2\Phi}={M\over\lambda}$, if 
${M\over\lambda}>{N\over12}$.\\\\
We now define the notion of apparent horizon.\\\\
In the four dimensional
theory, it is a hypersurface on which
${\left(\frac{\partial}{\partial r}\right)}^a$ (or equivalently $\nabla_ar$)
goes from being spacelike to timelike, i.e.\,, it is defined by the condition
${\left(\nabla r\right)}^2=0$. Here, the "radius" $r$ is the quantity which
gives the area $A$ of the two-spheres, i.e.\,, $r=\sqrt{{A\over4\pi}}$.\\\\
For a static black hole, the apparent horizon
coincides with the event horizon.\\\\
From the point of view of the two dimensional theory, this radius was
found to be proportional to $e^{-2\Phi}$; therefore, the apparent horizon is
where ${\left(\nabla\Phi\right)}^2=0$.\\\\
In the present case, the apparent horizon forms when 
\[\partial_+\Phi\left(x^+_0,x^-_0\right)=\Sigma\left(x^-_0\right)=0\;,\]
or, from~(\ref{eq:Sigma})    
\[\begin{array}{ll}
\ &1={M\over\lambda}\frac{e^{2\Phi}}{\sqrt{1-{N\over12}e^{2\Phi}}}\\\\
\Leftrightarrow&e^{-2\Phi}={N\over24}+\sqrt{{\left({N\over24}\right)}^2+
{\left({M\over\lambda}\right)}^2}\approx{M\over\lambda}+{N\over24}\\\\
\Leftrightarrow&x^-_0\approx-{M\over{\lambda^3x^+_0}}-{N\over{24\lambda^2x^+_0}}
\;,
\end{array}\]\\
where we have assumed that ${M\over\lambda}\gg N$.\\\\
The latter condition means that
the quantum corrections are very small where the apparent horizon forms:
\[Ne^{2\Phi}\approx N{\lambda\over M}\ll1\;.\]\\\\
In this case, we can find the slope of the apparent horizon
$\hat{x}^-\left(x^+\right)$ as follows:
\[0=\frac{d}{dx^+}{\left.\partial_+\Phi\right|}_{x^-=\hat{x}^-}=
\partial_+^2\Phi+\frac{d\hat{x}^-}{dx^+}\partial_+\partial_-\Phi\;,\]\\
because by definition $\partial_+\Phi=0$ on the apparent horizon.\\
Using (\ref{eq:BR2}), we get
\begin{eqnarray}
\label{eq:slope}
\frac{d\hat{x}^-}{dx^+}&=&\frac
{\left(-2\right)\left(1-{1\over12}Ne^{2\Phi}\right)\partial_+^{\,2}\Phi}
{\left(1-{1\over24}Ne^{2\Phi}\right)\lambda^2e^{2\rho}}\nonumber\\
\ &=&-\frac{e^{2\Phi-2\rho}\left(1-{1\over12}Ne^{2\Phi}\right)}
{\lambda^2\left(1-{1\over24}Ne^{2\Phi}\right)}\,T_{++}\;,
\end{eqnarray}\\
where we also have used the constraint equation for $g_{++}$ in the last line.
\\\\
Classically, $T_{++}$ is zero everywhere except at $x^+_0$, and the horizon is a
light-like line as expected.\\\\
When the quantum fluctuations of the matter fields
are taken into account, $T_{++}$ was found to be 
\[T_{++}=-{N\over24}\left(e^{2\rho}\,\partial_+^{\,2}e^{-2\rho}-
{1\over2}\,e^{4\rho}{\left(\partial_+e^{-2\rho}\right)}^2-2t_+\right)\;.\]\\\\
As long as $e^{-2\Phi}\gg N$, we can use the classical black hole solution to
evaluate the right hand side of~(\ref{eq:slope}). Then 
\[e^{-2\rho}\approx
e^{-2\Phi}\approx{M\over\lambda}-\lambda^2x^+\left(x^-+\Delta\right)\;,\]
and
\[\frac{d\hat{x}^-}{dx^+}\approx\left(-{1\over\lambda^2}\right)
\left(-{N\over24}\right)\left(-2t_+\right)\;.\]\\\\
$t_+$ is determined as before by the boundary condition that there is no 
incoming radiation from $J^-_R$:
\[\begin{array}{ll}
\ &0=T_{++}\left(x^+,x^-\rightarrow-\infty\right)=-{N\over24}
\left(0-{1\over2}{\left(\frac{-\lambda^2x^-}
{-\lambda^2x^+x^-}\right)}^2-2t_+\right)\\
\Rightarrow&t_+\left(x^+\right)=-{1\over4}{\left({1\over x^+}\right)}^2\;.
\end{array}\]\\\\
Thus 
\begin{equation}
\label{eq:aslope}
\frac{d\hat{x}^-}{dx^+}\approx\frac{N}{48\lambda^2}
{\left({1\over x^+}\right)}^2\;.
\end{equation}\\\\
Since the value of $e^{-2\Phi}$ on the apparent horizon is the mass
$m\left(x^+\right)$ of the evaporating black hole, this equation is valid as
long as $\frac{m\left(x^+\right)}{\lambda}\gg N$.\\\\
When $\frac{m\left(x^+\right)}{\lambda}\sim N$, the back reaction becomes
important and we can no longer use the classical black hole solution in
evaluating the right hand side of~(\ref{eq:slope}), but then the black hole
has radiated most of its mass because
\[\frac{m\left(x^+_0\right)}{\lambda}={M\over\lambda}\gg N\;.\]\\\\
Integrating~(\ref{eq:aslope}), we obtain
\[\hat{x}^-+\Delta_q=-{N\over{48\lambda^2}}{1\over x^+}\;.\]\\\\
As $x^+\rightarrow+\infty$, the apparent horizon thus approaches a global
horizon at 
\[\hat{x}^-=-\Delta_q=x^-_0+{N\over{48\lambda^2x^+_0}}=
-\Delta-{N\over{48\lambda^2x^+_0}}\;.\]\\\\
We can now show that this recession of the apparent horizon corresponds
precisely to the Hawking flux
\[\begin{array}{lll}
-dm\left(\hat{x}^-\right)&=&-d\left(\lambda e^{-2\Phi}\right)=
-\partial_+\left(\lambda e^{-2\Phi}\right)dx^+-
\partial_-\left(\lambda e^{-2\Phi}\right)dx^-\\
&=&-\partial_-\left(\lambda e^{-2\Phi}\right)dx^-\;.
\end{array}\]\\\\
In terms of the asymptotically minkowskian coordinates
\[\left\{\begin{array}{ccc}
\lambda x^+&=&e^{\lambda\sigma^+}\\
\lambda\left(x^-+\Delta_q\right)&=&-e^{-\lambda\sigma^-}
\end{array}\ \ \ \ ,\right.\]
\[\begin{array}{lll}
d\left(\lambda x^-\right)&=&e^{-\lambda\sigma^-}\,d\left(\lambda\sigma^-\right)=
-\lambda\left(x^-+\Delta_q\right)\,d\left(\lambda\sigma^-\right)\\\\
&=&{N\over{48\lambda x^+}}\,d\left(\lambda\sigma^-\right)\;.
\end{array}\]\\\\
Furthermore
\[\partial_-e^{-2\Phi}\approx-\lambda^2x^+\;.\]\\\\
So
\[-dm\left(\sigma^-\right)=\lambda\left(\lambda x^+\right)
{N\over{48\lambda x^+}}\,d\left(\lambda\sigma^-\right)=
{{N\lambda^2}\over48}\,d\sigma^-\;.\]\\\\
Thus the black hole looses mass at the same rate as it Hawking radiates
which is intuitively obvious but very difficult to show for e.g.\,, a
four dimensional Schwarzschild black hole.\\\\
So far, we have only taken into account the quantum corrections coming from
the conformal anomaly of the matter fields. In the large $N$ limit, they
dominate over the quantum
corrections of the dilaton field and of the conformal factor, but one would like
to consider a more systematic quantization of the classical CGHS model defined
by the action $S_{CGHS}$ given by (\ref{eq:CGHSa}). The corresponding quantum
theory is formally defined by the path integral
\[Z=\int \cal{D}\left(g,\Phi,f_i\right)e^{iS_{CGHS}}\;.\]\\\\
The general coordinate invariance allows us to gauge-fix the metric so that it
takes the form
\[g_{ab}=e^{2\rho}\hat{g}_{ab}\;,\]\\
where $\hat{g}_{ab}$ is a given reference metric which can be chosen to be the
Minkowski metric.\\\\
The gauge fixing gives rise to Fadeev-Popov ghosts, hence, the original path
integral over $g$ is replaced by an integral over the conformal factor $\rho$
and over the ghosts.\\\\
To define the theory, we must find the dependence of the measures in the path
integral on $\rho$. This yields a renormalized or effective action, the
renormalization depending on $\rho$ and also on $\Phi$ since $e^{2\Phi}$ is the
coupling strength of the theory.\\\\
The effective action should satisfy the following physical requirements:\\\\
First, it should not depend on the arbitrarily chosen reference metric
$\hat{g}$.
This can be achieved by requiring the action to be  both diffeomorphism
invariant with respect to $\hat{g}$ and conformally invariant, that is,
invariant under the transformation $\hat{g}\rightarrow e^{2\alpha}\hat{g}$. The
latter condition is equivalent to imposing that the total central $c$ vanishes.
\\\\
Second, as $e^{2\Phi}\rightarrow0$, the action should reduce to the classical
CGHS action and in addition, the leading order corrections in powers of
$e^{2\Phi}$ should describe a radiating black hole with a radiation flux
proportional to the number $N$ of matter fields.\\\\
Moreover, it is desirable that the semiclassical equations are solvable.\\\\
A more detailed review on the quantization of the CGHS model is given in
\cite[Section 4]{Th}, see also \cite{dA1,dA2,dA3}.\\\\
Several models satisfying the above requirements have been found
\cite{dA1,dA2,dA3,Bilal,RST1,RST2}, but here, we
will only describe the one proposed by Russo, Susskind and Thorlacius
\cite{RST1,RST2}.\\\\
These authors included a second one-loop term in addition to the Polyakov action
in a way that preserves a global symmetry present in the classical
CGHS action:
\[S=S_{CGHS}-{\kappa\over{8\pi}}\int d^2x\,\sqrt{-g}\,
\left(R\,{\left(\nabla^2\right)}^{-1}R+2\,\Phi\,R\right)\;,\]
$S_{CGHS}$ is the classical CGHS action, and $\kappa={N\over12}$.\\\\
One should also add the contributions from the ghosts, $\rho$ and $\Phi$, 
to the conformal anomaly but we will not elaborate this further. We only state
that it is possible to do this in such a way that $c=0$ and without affecting
our final results, see \cite{Lib}.\\\\
In the conformal gauge, the action is
\[\begin{array}{lll}
S&=&{1\over\pi}\int d^2x\left(e^{-2\Phi}\left(2\,\partial_+\partial_-\rho-
4\,\partial_+\Phi\partial_-\Phi+\lambda^2e^{2\rho}\right)+{1\over2}\,
\partial_+f_i\partial_-f_i\right)\\\\
&-&{\kappa\over\pi}\int d^2x\left(\partial_+\rho\partial_-\rho+
\Phi\,\partial_+\partial_-\rho\right)\;.
\end{array}\]\\\\
The constraints obtained from this action are 
\[\begin{array}{lll}
0&=&\left(e^{-2\Phi}+{\kappa\over4}\right)
\left(4\,\partial_{\pm}\rho\partial_{\pm}\Phi-2\,\partial_{\pm}^{\,2}\Phi\right)
+{1\over2}\,\partial_{\pm}f_i\partial_{\pm}f_i\\\\
&-&\kappa\left(\partial_{\pm}\rho\partial_{\pm}\rho-\partial_{\pm}^{\,2}\rho-
t_{\pm}\right)\;,
\end{array}\]
and the conserved current corresponding to the global symmetry is
\[j_a=\partial_a\left(\rho-\Phi\right)\;,\]
with
\[\partial^{\,a} j_a=\partial^{\,a}\partial_a\left(\rho-\Phi\right)=0\;,\]\\
as in the classical theory, we can choose coordinates, called the Kruskal
coordinates, in which $\rho=\Phi$.\\\\
The action can be rewritten as
\[S={1\over\pi}\int d^2x
\left(\partial_+\left(\kappa\rho+2e^{-2\Phi}\right)
\partial_-\left(\Phi-\rho\right)+\lambda^2e^{2\left(\rho-\Phi\right)}+
{1\over2}\,\partial_+f_i\,\partial_-f_i\right)\;.\]\\\\
This form suggests the definitions
\[\left\{\begin{array}{ccc}
\kappa\,\rho+2e^{-2\Phi}&=&\sqrt{\kappa}\left(\Omega+\chi\right)\\\\
\Phi-\rho&=&{1\over\sqrt{\kappa}}\left(\Omega-\chi\right)\;,
\end{array}\right.\]
or
\[\left\{\begin{array}{ccl}
\Omega&=&{\sqrt{\kappa}\over2}\,\Phi+\frac{e^{-2\Phi}}{\sqrt{\kappa}}\\\\
\chi&=&\sqrt{\kappa}\,\rho-{\sqrt{\kappa}\over2}\,\Phi+
\frac{e^{-2\Phi}}{\sqrt{\kappa}}\;.
\end{array}\right.\]\\\\
So
\[S={1\over\pi}\int d^2x
\left(-\partial_+\left(\chi-\Omega\right)\partial_-\left(\chi+\Omega\right)+
\lambda^2e^{{2\over\sqrt{\kappa}}\left(\chi-\Omega\right)}+
{1\over2}\,\partial_+f_i\,\partial_-f_i\right)\;,\]
and the constraints become
\[-\kappa t_{\pm}=-\partial_{\pm}\left(\chi-\Omega\right)
\partial_{\pm}\left(\chi+\Omega\right)+\sqrt{\kappa}\,\partial_{\pm}^{\,2}\chi+
{1\over2}\,\partial_{\pm}f_i\,\partial_{\pm}f_i\;.\]\\\\
The equations of motion for $\left(\chi-\Omega\right)$ resp.
$\left(\chi+\Omega\right)$ are
\[\left\{\begin{array}{lll}
-\partial_-\partial_+\left(\chi-\Omega\right)&=&0\\\\
-\partial_+\partial_-\left(\chi+\Omega\right)&=&\frac{2\lambda^2}{\sqrt{\kappa}}
e^{{2\over\sqrt{\kappa}}\left(\chi-\Omega\right)}\;.
\end{array}\right.\]\\\\
In Kruskal coordinates
\[\Omega-\chi=\sqrt{\kappa}\left(\Phi-\rho\right)=0\;,\]
and
\[\partial_+\partial_-\Omega=
\partial_+\partial_-\chi=-\frac{\lambda^2}{\sqrt{\kappa}}\;.\]\\\\
The solutions describing asymptotically flat static geometries are
\begin{equation}
\Omega=\chi=\frac{-\lambda^2x^+x^-}{\sqrt{\kappa}}+
P\,\sqrt{\kappa}\,\ln{\left(-\lambda^2x^+x^-\right)}+
\frac{M}{\lambda\sqrt{\kappa}}\;.
\end{equation}\\\\
The linear dilaton vacuum is still an exact solution with $P=-{1\over4}$ and
$M=0$.\\\\
The solution describing a collapsing shell of matter is
\begin{eqnarray}
\Omega&=&\chi=\frac{-\lambda^2x^+x^-}{\sqrt{\kappa}}-
\frac{\sqrt{\kappa}}{4}\,\ln{\left(-\lambda^2x^+x^-\right)}\nonumber\\
&-&\frac{M}{\lambda\sqrt{\kappa}x^+_0}\left(x^+-x^+_0\right)
\Theta\left(x^+-x^+_0\right)\;.
\end{eqnarray}\\\\
As is shown in \cite{RST1}, a curvature singularity forms on the infall line
when
\[\begin{array}{ll}
&\Omega'\left(\Phi\right)={\sqrt{\kappa}\over2}-{2\over\sqrt{\kappa}}\,
e^{-2\Phi}=0\\\\
\Leftrightarrow&e^{-2\Phi}={\kappa\over4},\ \ \ \Omega={\sqrt{\kappa}\over4}
\left(1-\ln{{\kappa\over4}}\right)\;.
\end{array}\]\\\\
For $x^+>x^+_0$, the curvature singularity lies on the critical line
$\Omega=\Omega_{cr}$ defined by
\[{\sqrt{\kappa}\over4}\left(1-\ln{{\kappa\over4}}\right)=
\frac{-\lambda^2x^+\left(x^-+\Delta\right)}{\sqrt{\kappa}}-
\frac{\sqrt{\kappa}}{4}\,\ln{\left(-\lambda^2x^+x^-\right)}+
\frac{M}{\lambda\sqrt{\kappa}}\;.\]\\\\
The critical line initially lies behind an apparent horizon on which
\[0=\partial_+\Omega\left(=\Omega'\,\partial_+\Phi\right)=
-{\lambda^2\over\sqrt{\kappa}}\left(x^-+\Delta\right)-
{\sqrt{\kappa}\over4}{1\over x^+}\;,\]
or
\[-\lambda^2\,x^+\left(x^-+\Delta\right)={\kappa\over4}\;.\]\\\\
Therefore, the apparent horizon recedes at a rate 
\[\frac{dx^-}{dx^+}=\frac{\kappa}{4\lambda^2{\left(x^+\right)}^2}\;.\]\\
As in the original CGHS model, this rate of recession corresponds to the 
Hawking flux.\\\\
The critical line and the apparent horizon
\[\left\{\begin{array}{ccl}
{\kappa\over4}\left(1-\ln{{\kappa\over4}}\right)&=&
-\lambda^2x^+\left(x^-+\Delta\right)-
{\kappa\over4}\,\ln{\left(-\lambda^2x^+x^-\right)}+
\frac{M}{\lambda}\\\\
-\lambda^2x^+\left(x^-+\Delta\right)&=&{\kappa\over4}\;,
\end{array}\right.\]
meet when
\[\left\{\begin{array}{ccc}
-\lambda^2x^+\left(x^-+\Delta\right)&=&{\kappa\over4}\\\\
-\lambda^2x^+x^-&=&{\kappa\over4}\,e^{\frac{4M}{\kappa\lambda}}\;.
\end{array}\right.\]\\\\
Thus
\[\left\{\begin{array}{lll}
x^+_s&=&x^+_0\,\frac{\kappa\lambda}{4M}\,\left(e^{\frac{4M}{\kappa\lambda}}
-1\right)\\\\
x^-_s&=&-\frac{\Delta}{1-e^{-\frac{4M}{\kappa\lambda}}}\;.
\end{array}\right.\]\\\\
At this point, the curvature on the apparent horizon is infinite: on the 
apparent horizon, the curvature is given by
\[\begin{array}{lll}
R=\frac{4\lambda^2}{\left(1-{\kappa\over4}\,e^{2\Phi}\right)}\rightarrow+\infty
&as&e^{-2\Phi}\rightarrow e^{-2\Phi_{cr}}={\kappa\over4}\;.
\end{array}\]\\\\
The critical line, which was spacelike before it met the apparent horizon
(since it was behind the apparent horizon), becomes timelike (now it is in
front of the apparent horizon).\\\\
The spacetime diagram is shown in fig. \ref{fig:RST}.
\begin{figure}[t]
	\vspace{7 cm}
	\includegraphics{figbh/RST.eps}
	\caption{}
	\label{fig:RST}
\end{figure}
\\\\It can be shown \cite{RST1,RST2} that if we modify $\Omega$ in region $III$
by adding a
function of $x^-$ (so that the $g_{++}$-constraints still is satisfied) in such
a way
that the curves $\Omega=\Omega_{cr}$ and $\partial_+\Omega=0$ merge, i.e.\,, we
impose the boundary condition
${\left.\partial_+\Omega\right|}_{\Omega=\Omega_{cr}}=0$ (which also implies
${\left.\partial_-\Omega\right|}_{\Omega=\Omega_{cr}}=0$), then the curvature
will be finite on the critical line which can thus be viewed as the boundary of
spacetime (analogous to the boundary $r=0$).\\\\
In region $III$, the modified solution
turns out to describe a shifted dilaton vacuum:
\[e^{-2\Phi}=e^{-2\rho}=-\lambda^2x^+\left(x^-+\Delta\right)\;.\]\\\\
Now
\[\begin{array}{cl}
\Omega&=-\frac{\lambda^2}{\sqrt{\kappa}}\,x^+\left(x^-_s+\Delta\right)-
{\sqrt{\kappa}\over4}\ln{\left(-\lambda^2x^+x^-_s\right)}+
{M\over{\lambda\sqrt{\kappa}}}\\\\
\ &=-\frac{\lambda^2}{\sqrt{\kappa}}\,x^+\left(x^-_s+\Delta\right)-
{\sqrt{\kappa}\over4}
\left(\ln{\left(-\lambda^2x^+x^-_s\right)}-\frac{4M}{\lambda\kappa}\right)\\\\
\ &=-\frac{\lambda^2}{\sqrt{\kappa}}\,x^+\left(x^-_s+\Delta\right)-
{\sqrt{\kappa}\over4}\ln{\left(-\lambda^2x^+\left(x^-_s+\Delta\right)\right)}\;,
\end{array}\]
where we used
\[1+{\Delta\over x^-_s}=e^{-\frac{4M}{\kappa\lambda}}\;.\]\\\\
Therefore, the solution is continuous across $x^-=x^-_s$. Evaluating
$\partial_-\Omega$, we find that its change across $x^-=x^-_s$ is
\[-{\sqrt{\kappa}\over4}\left({1\over{x^-_s+\Delta}}-{1\over x^-_s}\right)=
{\sqrt{\kappa}\over4}\frac{\Delta}{x^-_s\left(x^-_s+\Delta\right)}\;.\] \\\\  
This corresponds to a shock wave
\[\begin{array}{cl}
T_{--}^f&={1\over2}\,\partial_- f_i\partial_- f_i=
-\sqrt{\kappa}\,\partial_-^{\,2}\Omega={\kappa\over4}\,
\frac{\Delta}{x^-_s\left(x^-_s+\Delta\right)}\,\delta\left(x^--x^-_s\right)\\\\
\ &=-{{\kappa\lambda}\over4}\left(1-e^{-\frac{4M}{\kappa\lambda}}\right)
{\left(-\lambda\left(x^-_s+\Delta\right)\right)}^{-1}
\delta\left(x^--x^-_s\right)\;.
\end{array}\]\\\\
This shock wave carries a small amount of negative energy
$-{{\kappa\lambda}\over4}\left(1-e^{-\frac{4M}{\kappa\lambda}}\right)$
out to infinity.\\\\
In this model therefore, the black hole evaporates completely.

\newpage
\section{Conclusions}
As we have seen in section 3, Hawking's calculation of the black hole radiance
yields radiation in a mixed thermal state. For a "superobserver", who is able
to see
the quantum state $\psi$ of the total Fock space $\cal{F}_{H^+}\otimes
\cal{F}_{J^+}$, the state is pure but from the point of view of an observer
outside the black hole, who can only access $\cal{F}_{J^+}$, the same state is
viewed as a mixed state.\\\\
Most of the information contained in the incoming state will end up behind
the event horizon of the black hole. This information which from the point of
view of the external observer is lost, can be quantified by calculating the
so-called entropy of entanglement
which is given by 
\[E=-tr{\,\rho\ln{\rho}}\;,\]
where $\rho$ is the density matrix describing the mixed thermal state at future
null infinity. If $\rho$ had arised from a pure state, the entropy would have
been zero but in the case of a mixed state, it is positive.\\\\
As we saw in
section 3, the density matrix is obtained from $\psi$ by "tracing over" all
horizon states
but it can also be obtained directly by applying the so-called superscattering
operator $\$$ on the in-state; $\$$ is obtained from the $S$-matrix by tracing 
over the horizon states
\[\$=tr_{H^+}\,S\,S^{\dagger}\;.\]\\\\
We recall that the calculation of $\rho$ is valid only in the semiclassical
regime, when the gravitational field of the collapsing star can be described by
a classical real-valued metric. Therefore, the calculation breaks down when the
mass of the black hole approaches the Planck mass because at this point, the
quantum fluctuations of the metric are expected to become important and of
course, the theory describing the quantum fluctuations at this scale is
unknown.\\\\
We assume for the moment that the unknown planckian dynamics cause the black
hole to evaporate completely {\em and} that the information disappears with the
black hole, i.e.\,, the out-state is still described by a mixed state even
after the evaporation. In this case, the evolution from in-state to out-state
is governed by the superscattering operator $\$$ rather than by a unitary
$S$-matrix; in other words, the usual rules of quantum mechanics are modified
in the presence of black holes. This very radical proposal was made by Hawking
in 1976 \cite{Hawk76}. Since then, however, several authors have argued that
this $\$$-matrix
evolution would violate energy conservation and destabilize the vacuum, see
\cite{Banks,St'HW}.\\\\
Another possibility is that the planckian physics shuts off the Hawking
radiation when the black hole reaches the Planck mass. Hence, the information
is stored forever in a Planck-mass remnant, and quantum mechanics is not
violated. The main criticism against this scenario is that it is difficult to
believe in the stability of the remnant in the absence of any conservation law.
\\\\
A third proposal is that the information is encoded in the Hawking radiation
which therefore is in a pure state.\\\\
One possibility is that the information is reemitted during the whole lifetime
of the black hole, in direct contradiction with Hawking's calculation. Naively, it
seems impossible to realize this idea because the information cannot be
reemitted to future null infinity and at the same time go to the black hole:
if we assume that a pure in-state evolves to an out-state which is a product
state, then a superposition of in-states would evolve to a mixed state; hence,
the assumption that the out-state is always a product state is not consistent
with the superposition principle. One says that information duplication is not
possible in quantum mechanics.\\\\
However, the above argument against unitary evolution has been criticized by
Susskind who postulates that there is no combined quantum description for a
freely falling observer approaching the horizon and a distant observer who
remains outside the black hole. Quantum mechanics is valid for each of them
separately and in particular, for the external observer, the evolution of
quantum states is unitary. Contradictions would certainly arise if the external
observer and the freely falling observer could compare their experiments but
they cannot, since the freely falling observer disappears into the black hole;
the two descriptions are {\em complementary}. This principle of black hole
complementarity is the basis of the approach by Susskind \cite{Suss1,Suss3}.
This principle implies that Planck scale physics enter
in the description of the distant observer during the whole process of the
evaporation
of the black hole, not just in its final stage, as is assumed in Hawking's
calculation. This is because the redshift between a point
close to the horizon and infinity is enormous.
\\\\
Susskind has proposed to describe a black hole in terms of a membrane just
outside
the event horizon (its area is defined to be one Planck unit larger than the
area of the event horizon), the so-called stretched horizon, which would absorb
and reemit all the information contained in the incoming state \cite{Suss1}.
The membrane description had already been useful for
classical black holes, see the references in \cite{Suss1}. From the point of 
of view of the distant observer, the stretched horizon behaves like a real
physical membrane. In particular, its temperature is so high (of the order of
the Planck temperature), that a freely falling observer is seen to be destroyed;
but
the freely falling observer sees himself passing through the event horizon
without problem, since from his point of view, the stretched horizon does not
even exist!\\\\
Susskind \cite{Suss2,Suss4}, and before him 't Hooft \cite{'t Hooft90},
have speculated that the degrees of freedoom
of the stretched horizon are described by string theory but it is not known if
string theory really provides a mechanism by which the stretched horizon can
reemit the original information encrypted in the outgoing radiation, although
there are some indications in favor of such a possibility.\\\\
Still another possibility is that the information is reemitted after the black
hole mass has reached the Planck mass; the restoration of information is then
governed by planckian dynamics which we know nothing about. However, general
arguments indicate that the information is reradiated very slowly \cite{Presk}:
a four
dimensional black hole with initial mass $M$ evaporates down to Planck size in
time $M^3$, but the decay time for the Planck-size remnant is much longer, at
least $M^4$, which for a mass of the order of the sun mass is an enormously
large number ($10^{108}\,s\;!?!\,$).
The problem is again how to explain such a very long decay time,
so we are back to the case of an absolutely stable remnant.
\\\\
Have the two dimensional models resolved the information paradox?\\\\
In the RST model, the information does not come back, at least as long as the
semiclassical equations are valid. Hence, if the information comes back at all,
it is when the black hole reaches the analog of the Planck mass and the result
would be a long-lived remnant. Unfortunately, there are other versions of two
dimensional gravity which are claimed not to lead to information loss
\cite{Verl1,Verl2} so even
in two dimensions, the information paradox has not been resolved. In fact, the
physical relevance of two dimensional models has been questionned: it seems
likely that it is not possible to truncate the four dimensional theory to
include only spherically symmetric modes \cite{SU}.\\\\
One possible resolution of the black hole information paradox is simply that
black holes do not exist. One should perhaps look for alternatives to general
relativity, for example the recently proposed
non-symmetric gravitational theory \cite{Moff}. The static spherically
symmetric solution
of this theory, corresponding to the Schwarzschild solution, is everywhere
regular and does not contain any event horizon. Black holes are replaced by
superdense objects that are stable for arbitrary large masses, due to a new
repulsive force that counterbalances the usual attractive gravitational force.
Moreover, by choosing a particular parameter entering in the solution to be
sufficiently small, the theory agrees with general relativity for the scales
where it offers a good description of the experimental data.\\\\\\\\
Even if this new theory does not survive, it remains an open possibility
that
black holes do not exist, because they still have to be identified with
certainty.


\newpage
\begin{thebibliography}{99}

\bibitem{Wald}
R. M. Wald, {\it General Relativity}, The University of Chicago Press, 1984.

\bibitem{grav}
C. Misner, K. Thorne, and J. Wheeler, {\it Gravitation}, W. H. Freeman, New
York, 1973.

\bibitem{Hawk75}
S. W. Hawking, {\it Particle Creation by Black Holes}, Comm. Math. Phys.
{\bf43} (1975) 199.

\bibitem{Wald75}
R. M. Wald, {\it On Particle Creation by Black Holes}, Comm. Math. Phys.
{\bf45} (1975) 9.

\bibitem{Hawk76}
S. W. Hawking, {\it Breakdown of Predictability in Gravitational Collapse},
Phys. Rev. {\bf D 14} (1976) 2460.

\bibitem{Birr}
N. D. Birrell and P. C. W. Davies, {\it Quantum Fields in Curved Space},
Cambridge University Press, 1982.

\bibitem{GSW}
M. Green, J. H. Schwarz and E. Witten, {\it Superstring Theory, Vol. 1},
Cambridge University Press, 1987.

\bibitem{GHS}
D. Garfinkle, G. T. Horowitz, A. Strominger, {\it Charged Black Holes in String
Theory}, Phys. Rev. {\bf D 43} (1991) 3140; Erratum,
Phys. Rev. {\bf D 45} (1992) 3888.

\bibitem{CGHS}
C. G. Callan, S. B. Giddings, J. A. Harvey, A. Strominger, {\it Evanescent
Black Holes}, Phys. Rev. {\bf D 45} (1992) R 1005.

\bibitem{Wit}
E. Witten, {\it String Theory and Black Holes}, Phys. Rev. {\bf D 44} (1991)
314.

\bibitem{Cardy}
J. L. Cardy, {\it Conformal Invariance and Statistical Mechanics}, {\it in}
Les Houches 1988, North Holland, 1990.

\bibitem{Alv}
L. Alvarez-Gaum\'{e} and E. Witten, {\it Gravitational Anomalies},
Nucl. Phys. {\bf B 234} (1983) 269.

\bibitem{Pol}
A. M. Polyakov, {\it Quantum Geometry of Bosonic Strings},
Phys. Lett. {\bf 103 B} (1981) 207.

\bibitem{GN}
S. B. Giddings and W. M. Nelson, {\it Quantum Emission from Two-Dimensional
Black Holes}, Phys. Rev. {\bf D 46} (1992) 2486.

\bibitem{Ginsparg}
P. Ginsparg, {\it Applied Conformal Field Theory}, {\it in}
Les Houches 1988, North Holland, 1990.

\bibitem{ST1}
J. G. Russo, L. Susskind and L. Thorlacius, {\it Black Hole Evaporation in 1+1
Dimensions}, Phys. Lett. {\bf B 292} (1992) 13.

\bibitem{ST2}
L. Susskind and L. Thorlacius, {\it Hawking Radiation and Back-Reaction},
Nucl. Phys. {\bf B 382} (1992) 123.

\bibitem{Th}
L. Thorlacius, {\it Black Hole Evolution}, Santa Barbara preprint
NSF-ITP-94-109, hep-th/9411020.

\bibitem{dA1}
S. P. de Alwis, {\it Quantization of a Theory of 2D Dilaton Gravity}, 
Phys. Lett. {\bf B 289} (1992) 278.

\bibitem{dA2}
S. P. de Alwis, {\it Black Hole Physics from Liouville Theory},
Phys. Lett. {\bf B 300} (1993) 330.

\bibitem{dA3}
S. P. de Alwis, {\it Quantum Black Holes in Two Dimensions},
Phys. Rev. {\bf D 46} (1992) 5429.

\bibitem{Bilal}
A. Bilal and C. Callan, {\it Liouville Models of Black Hole Evaporation},
Nucl. Phys. {\bf B 394} (1993) 73.

\bibitem{RST1}
J. G. Russo, L. Susskind and L. Thorlacius, {\it End Point of Hawking
Radiation}, Phys. Rev. {\bf D 46} (1992) 3444.

\bibitem{RST2}
J. G. Russo, L. Susskind and L. Thorlacius, {\it Cosmic Censorship in
Two-Dimensional Gravity}, Phys. Rev. {\bf D 47} (1993) 533.

\bibitem{Lib}
S. Liberati, {\it A Real Decoupling Ghost Quantization of the CGHS Model for
Two-Dimensional Black Holes}, hep-th/9407002.
 
\bibitem{Banks}
T. Banks, L. Susskind and M. E. Peskin, {\it Difficulties for the Evolution of
Pure States into Mixed States}, Nucl. Phys. {\bf B 244} (1984) 125.

\bibitem{St'HW}
C. R. Stephens, G. 't Hooft, B. F. Whiting, {\it Black Hole Evaporation without
Information Loss}, Class. Quant. Grav. {\bf 11} (1994) 621.

\bibitem{Suss1}
L. Susskind, L. Thorlacius and J. Uglum, {\it The Stretched Horizon and Black
Hole Complementarity}, Phys. Rev. {\bf D 48} (1993) 3743.

\bibitem{Suss3}
L. Susskind and L. Thorlacius, {\it Gedanken Experiments Involving Black
Holes}, Phys. Rev. {\bf D 49} (1994) 966.

\bibitem{Suss2}
L. Susskind, {\it String Theory and the Principle of Black Hole
Complementarity}, Stanford preprint SU-ITP-93-18, July 1993, hep-th/9307168.

\bibitem{Suss4}
L. Susskind, {\it Strings, Black Holes and Lorentz Contraction},
Stanford preprint SU-ITP-93-21, August 1993, hep-th/9308139.
 
\bibitem{'t Hooft90}
G. 't Hooft, {\it The Black Hole Interpretation of String Theory},
Nucl. Phys. {\bf B 335} (1990) 138.

\bibitem{Presk}
J. Preskill, {\it Do Black Holes Destroy Information ?}, Caltec preprint
CALT-68-1819, hep-th/9209058.

\bibitem{Verl1}
E. Verlinde and H. Verlinde, {\it A Quantum S-matrix for Two-Dimensional
Black Hole Formation and Evaporation}, Nucl. Phys. {\bf B 406} (1993) 43.

\bibitem{Verl2}
K. Schoutens, E. Verlinde and H. Verlinde, {\it Quantum Black Hole
Evaporation}, Phys. Rev. {\bf D 48} (1993) 2670.

\bibitem{SU}
L. Susskind and J. Uglum, {\it Black Holes, Interactions, and Strings},
Stanford preprint SU-ITP-94-35, hep-th/9410074.

\bibitem{Moff}
N.J. Cornish and J. W. Moffat, {\it Non-Singular Gravity Without Black Holes},
Toronto preprint UTPT-94-08, gr-qc/9406007.

\end{thebibliography}
\end{document}